\documentclass{article}

\usepackage[main, final]{neurips_2026}
\usepackage{amsmath}
\usepackage{tikz}
\usepackage[most]{tcolorbox}

\usepackage[utf8]{inputenc}
\usepackage[T1]{fontenc}
\usepackage{hyperref}
\usepackage{url}
\usepackage{booktabs}
\usepackage{amsfonts}
\usepackage{nicefrac}
\usepackage{microtype}
\usepackage{xcolor}
\usepackage{graphicx}

\usepackage{xcolor}

\definecolor{royalblue}{rgb}{0.21,0.49,0.74}

\usepackage{wrapfig}

\title{Controllable and Content-Based Recommendations}

\author{%
  \textbf{Fırat Öncel}$^{1,5}$ \quad
  \textbf{Jihoon Jeong}$^{3,5}$ \quad
  \textbf{Emiliano Penaloza}$^{2,5}$ \\
  \textbf{Mirco Ravanelli}$^{1,5}$ \quad
  \textbf{Laurent Charlin}$^{4,5}$ \quad
  \textbf{Cem Subakan}$^{1,3,5}$ \\
  \vspace{2pt} \\
  \normalfont
  $^{1}$Concordia University \quad
  $^{2}$Université de Montréal \quad
  $^{3}$Laval University \\
  $^{4}$HEC Montréal \quad 
  $^{5}$Mila -- Quebec AI Institute \\[4pt]
  \texttt{firat.oncel@mila.quebec} \quad
}

\begin{document}
\maketitle

\newcommand{\cem}[1]{{ \color{blue} $\mathcal C:$  #1}}
\newcommand{\firat}[1]{{ \color{red} $\mathcal F:$  #1}}

\begin{abstract}
Traditional recommendation systems rely on latent (dense) representations, making them difficult to interpret and control. We propose the Controllable and Content-Based Recommendations (CCBR) framework, which builds its recommendations from textual user profile representations. CCBR plugs into collaborative filtering models and introduces controllability via text bottlenecks. We show that CCBR enables text-based and multimodal interventions, allowing users to steer the model towards the directions they prefer. Different from existing controllable recommendation systems, CCBR infers the text summaries directly from item contents (images, audio or video). Across image, audio, and video datasets, we demonstrate that the proposed framework obtains competitive model performance with standard (latent-representation) models while providing controllable model summaries via text. The model also outperforms TEARS, a recent baseline for controllable recommendation systems. Through systematic interventions, we demonstrate the efficacy of the steering mechanism. 
\end{abstract}

\begin{tcolorbox}[colback=gray!8, colframe=gray!50, arc=2mm, boxrule=0.5pt]
\small
\textbf{Project Page.}
For additional demo materials, please visit:
\href{https://firatoncel.github.io/ccbr/}{\textcolor{royalblue}{https://firatoncel.github.io/ccbr/}}.

\textbf{Code.} To access project codes, please visit \href{https://github.com/firatoncel/CCBRNeurIPS2026}{\textcolor{royalblue}{https://github.com/firatoncel/CCBRNeurIPS2026}}
\end{tcolorbox}

\section{Introduction}
\label{sec:intro}

Recommendation systems play a central role in helping users navigate increasingly large catalogs of music, products, movies, and other digital content \cite{ricci2021recommender}. 
Despite their practical success, modern recommendation systems often remain opaque to users \cite{abdollahi2016explainable, caro2021conceptual}.
Many state-of-the-art recommendation systems are based on dense architectures and operate via collaborative filtering \cite{he2017neural, ease}, which are effective for prediction but difficult to interpret \cite{almutairi2021xpl}. 
As a result, users have limited insight into why particular items are recommended and little direct control over the system, especially when their current intent differs from the preferences reflected in their past behavior.

This lack of controllability is particularly limiting in scenarios where preferences are context-dependent \cite{adomavicius2010context}. 
For example, a user shopping for clothing may want to temporarily emphasize color, texture, or aesthetic, rather than receive recommendations that simply reflect their long-term purchase history. 
Standard collaborative filtering methods are not designed to expose such semantic factors to users \cite{koren2009matrix, he2017neural, mysore2023editable}. 
Although they are scalable and well-grounded in the item catalog, their internal representations do not inherently provide a user-editable interface for controlling preferences.

Recent work has explored more transparent and controllable recommendation architectures, including systems that condition their final decisions on interpretable bottleneck layers \cite{positionScrutable, tears, oncel2025audioprototypicalnetworkcontrollable}.
These approaches suggest that exposing intermediate preference representations can make recommendation systems more interpretable and steerable. 
In many real-world domains, user preferences are expressed through rich multimodal content: songs, product images, videos, or other media. 
A controllable recommender should therefore be able to derive semantic user profiles directly from such content while still allowing users to inspect and edit those profiles in natural language.

In order to realize the full potential of text-based representations,
this work proposes Controllable and Content-Based Recommendations (CCBR), a content-based recommendation system that creates scrutable and controllable user summaries through item contents. 
These summaries aggregate user history in a transparent way, so that user preferences can be presented in natural language. The user can understand how the system represents them, and steer it towards the directions they would like the recommendations to take.  
Unlike earlier controllable recommendation systems \cite{positionScrutable, tears, oncel2025audioprototypicalnetworkcontrollable}, CCBR is content-based, deriving user profiles directly from images, audio, or video. 
Users can interact with the system through multimodal inputs and by editing its textual representations. 
For example, in a music recommendation scenario, users can upload some of their favorite songs (that the system has not previously seen) to help the model capture more subtle aspects of their taste in the resulting text summary. 
This is different than existing controllable recommendation systems, as the system ingests the multimodal content directly, and produces the summary. 
Alternatively, they can directly edit their text-based user summary (e.g., by adding relevant words or sentences) to steer the model toward their current musical preferences.

To achieve this, CCBR generates text-based user summaries by first encoding items in the user’s history with multimodal LLMs and then aggregating these item-level representations using a text-based LLM. 
The resulting summary is passed to a multi-label classification head that predicts the high-level concepts associated with the user. 
Finally, we use a latent representation guided by multi-label supervision to regularize collaborative filtering embeddings, making recommendation models more interpretable and controllable.
We evaluate CCBR across music, shopping, and movie recommendation tasks, demonstrating its ability to generate relevant user summaries. 
We also show that our method can be successfully applied on top of several state-of-the-art recommendation systems, including EASE \cite{ease}, EDLAE \cite{edlae}, DAE \cite{edlae}, Multi-DAE \cite{mult_dae_vae}, Multi-VAE \cite{mult_dae_vae}, and MacridVAE \cite{macridvae}, making them controllable.

Our contribution is as follows:
\begin{itemize}
    \item We propose CCBR, a framework that introduces controllability to a given collaborative filtering model in a content-driven, multimodal way through a text-based, editable user summary (Section~\ref{sec:method}).  
    \item We demonstrate that, despite adding controllability, CCBR does not significantly deteriorate the recommendation performance of several popular models, and outperforms a controllable recommendation system baseline (Section~\ref{sec:performance}).
    \item We show that CCBR can steer recommendations by incorporating previously unseen multimodal inputs across H\&M, MovieLens-20M (ML-20M), and Million Song Dataset (MSD) datasets (Sections~\ref{sec:controllability} and~\ref{sec:loi_unseen}).
\end{itemize}

\section{Related Works}

\paragraph{Collaborative Filtering-based Recommendation Systems.} 
Collaborative filtering (CF) is a foundational paradigm in recommender systems, which infers user preferences from historical user-item interactions \cite{ko2022survey}. 
CF methods have progressed from similarity-based approaches \cite{resnick1994grouplens, sarwar2001item, linden2003amazon} that rely on user-user or item-item neighborhoods, to linear reconstruction models such as EASE \cite{ease} and EDLAE \cite{edlae}, and further to neural or variational autoencoder-based models such as DAE \cite{edlae}, Multi-DAE \cite{mult_dae_vae}, Multi-VAE \cite{mult_dae_vae}, and MacridVAE \cite{macridvae}.
These methods remain effective in large-scale settings since they directly capture collaborative signals and enable efficient retrieval or ranking over a fixed item catalog.

However, conventional CF models primarily encode user preferences as latent, ID-based representations that are difficult to interpret, inspect, or modify \cite{balog2019transparent, mysore2023editable}. 
Existing content-aware CF methods can incorporate side information to improve recommendation accuracy, but such information is typically used internally by the model rather than exposed as an explicit user-editable control interface \cite{chen2020attribute}. 
This limits their applicability to controllable recommendations, in which users may want to directly adjust the semantic factors that influence their results. 
To address this gap, CCBR exposes user preferences through editable semantic concepts while preserving the scalability and catalog-grounding benefits of CF. 
By making preferences accessible through human-understandable concepts, CCBR aligns with the motivation of concept bottleneck models \cite{koh2020concept, shin2023closer}, which improve interpretability and enable intervention via explicit intermediate concepts.
\vspace{-5pt}
\paragraph{LLM-based Recommendation Systems.} 
Large language models (LLMs) have recently been explored as a way to incorporate rich textual semantics into recommender systems \cite{wu2024survey, bao2023tallrec}, providing a flexible text-based interface and producing human-readable reasoning. 
However, directly using LLMs as recommenders remains challenging in large-scale catalog settings, where recommendations must be consistent, grounded in available items, and efficiently retrieved from a fixed item set. 
Moreover, zero-shot and few-shot methods \cite{liu2023chatgpt, dai2023uncovering} can suffer from hallucination or incomplete catalog coverage, while fine-tuned LLM recommenders \cite{kang2023llms, zheng2024harnessing} often require task-specific adaptation and still need mechanisms for reliable catalog grounding.

Another line of work uses LLM-generated text to enhance existing recommendation models \cite{ wei2024llmrec, ren2024representation}, generating user- or item-level textual attributes and integrating them into conventional collaborative filtering. 
This can improve the quality of representation by exposing high-level semantic information that is not easily captured by sparse user-item interactions alone \cite{tears, lyu2024llm}. 
For example, TEARS \cite{tears} introduces a controllable system that summarizes user preferences in natural language, enabling users to inspect and edit the summaries. 
However, TEARS and most related approaches rely primarily on textual metadata, such as item titles, descriptions, or user-written content \cite{lyu2024llm, yin2023heterogeneous}.
As a result, they fail to capture preference-relevant signals that are expressed through non-textual content, such as visual style in products, acoustic characteristics in music, or temporal and aesthetic cues in videos.

Recent work has also explored adapting multimodal large language models (MLLMs) to recommendation \cite{liu2024rec, ye2025harnessing}. Rec-GPT4V \cite{liu2024rec} uses MLLMs to summarize item images and reason over user preferences, while MLLM-MSR \cite{ye2025harnessing} fine-tunes an MLLM-based recommender for sequential multimodal recommendation. Although these approaches demonstrate the potential of multimodal foundation models for recommendation, they mainly focus on improving recommendation accuracy or reasoning capability. The user preference representation is often implicit in the hidden states, making it difficult for users to directly control the recommendation process. Therefore, CCBR bridges multimodal content understanding and controllable collaborative filtering through a cascade that converts multimodal items into textual summaries and merges them into an editable user-level profile.

\vspace{-5pt}
\paragraph{Controllable and Preference-Driven Recommendation.}
Controllability has recently become an explicit design goal in both research and recommendation systems in production, with natural language serving as a common interface for it. A useful way to organize these approaches is by where language enters the recommendation pipeline. In narrative-driven recommendation \cite{mysore_llm}, an LLM is used offline to synthesize training queries for a retrieval model, so language serves as a query rather than as a learned user representation. A second group keeps the recommender's representation latent and introduces language alongside it: EXP3RT \cite{kim_review} adds an LLM reasoning layer that summarizes reviews to predict and explain ratings, CTRL-Rec \cite{carroll2025ctrlreccontrollingrecommendersystems} models how natural-language requests affect user-item affinities and uses this as a request-conditioned signal over an engagement score, and Mender \cite{paischer2025preference} conditions a generative retrieval model on a natural-language preference string at inference time. Beyond research systems, controllability has also begun to appear in production, where Instagram lets users adjust and reset the interests that drive their Reels recommendations \cite{instagram_reels_control}. CCBR is complementary to these efforts: rather than attaching language to a latent representation, it uses an interpretable text profile as the user representation itself, and CCBR derives the presented profile from raw item content across image, audio, and video rather than from reviews, metadata, or interaction history.

\section{Methodology}
\label{sec:method}

\begin{figure*}[t]
    \centering
    \includegraphics[trim=1cm 1cm 1cm 1cm, clip, width=.95\linewidth]{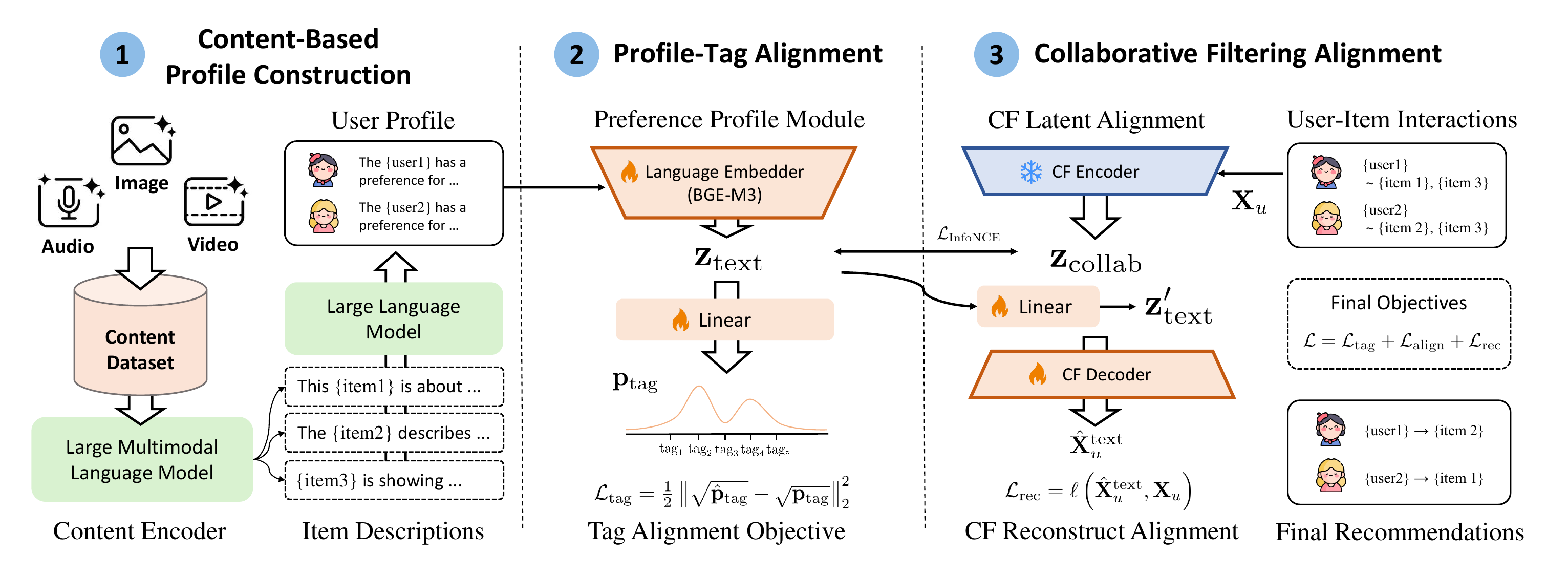}
\caption{\textbf{Overview of the CCBR pipeline.} 
1) Item-level content (images, audio, or video) is first verbalized into natural-language descriptions by a multimodal language model, and a language model then aggregates each user's interacted items into a textual preference profile. 2) The profile is mapped to a dense representation through a trainable text encoder and a linear projection head. 3) This text-side representation is aligned with the user representation produced by a frozen collaborative filtering backbone, whose decoder is reused to score candidate items.}
    \label{fig:ccbr_pipeline}
\end{figure*}

We introduce CCBR, a recommendation framework that enables controllability of a collaborative filtering model. In this framework, user preferences are represented as natural-language profiles derived from the content of items that the user interacted with. Unlike prior LLM-based recommendation systems \cite{bao2023tallrec, llara},  that incorporate items via item identifiers, CCBR enables the creation of user summaries through multimodal content ingestion.  
Specifically, CCBR generates user profiles from item \emph{content} (images, audio, or video) and uses the LLM as a summarization engine over the item summaries. The resulting text profile serves as a human-readable, editable representation of user taste, while final scoring is performed by a collaborative filtering (CF) backbone that is aligned with the text representation during training. The overall pipeline is given in Figure~\ref{fig:ccbr_pipeline}.

\vspace{-5pt}
\paragraph{(1) User profile generation.}
The raw content of each item is passed through a modality-specific multimodal foundation model to produce a concise natural-language description. This step is performed once per item, offline, and the resulting descriptions are cached and reused across all subsequent training and inference. For each user, the descriptions of items in their interaction history are then passed to an instruction-tuned text-only LLM, which produces a natural-language profile summarizing the user's preferences. Crucially, item identifiers such as titles, product IDs, or artist names are never exposed to either the multimodal or the text-only model, so the generated profile reflects only a content-derived signal, and no pretraining knowledge about specific items can leak into the user representation. The specific models used for each modality are detailed in Section~\ref{sec:experiments}.
\vspace{-5pt}
\paragraph{(2) Text encoding with tag-supervised pretraining.}
We encode the natural-language user profile with BGE-M3 \cite{bgem3}, producing a text representation $\mathbf{z}_{\text{text}} \in \mathbb{R}^{1024}$. 
While general-purpose text encoders capture broad semantic similarity, they are not optimized for the recommendation domain: two summaries that read alike at the surface level may correspond to very different taste profiles, and conversely, summaries phrased differently may describe the same underlying preferences. 
Dataset-specific tags, movie genres, product categories and colors, music tags, provide a compact and human-interpretable view of user taste that is directly tied to the catalog the CF backbone scores over. 
Supervising the text encoder with tag distributions therefore serves two purposes: 
(i) it grounds the textual representation in the same content vocabulary used to evaluate controllability in Section~\ref{sec:controllability}, ensuring that lexical edits to the profile translate into meaningful shifts along interpretable axes, and 
(ii) it injects a recommendation-aware inductive bias into the encoder prior to CF alignment, allowing the subsequent alignment stage to start from representations whose geometry already reflects item-content structure rather than generic textual similarity.  
We train the text encoder in two stages.

In the \emph{first stage}, we leverage dataset-specific tag annotations available for each item. For each user, we construct a target tag distribution $\mathbf{p}_{\text{tag}} \in \Delta^{n_{\text{tags}}}$ by aggregating the tags of items in their interaction history and normalizing to a valid probability distribution. A linear head projects the text embedding to a predicted tag distribution,
\begin{equation}
\hat{\mathbf{p}}_{\text{tag}} = \mathrm{softmax}(\mathbf{W}\, \mathbf{z}_{\text{text}} + \mathbf{b}),
\qquad \mathbf{W} \in \mathbb{R}^{n_{\text{tags}} \times d},
\end{equation}
and the encoder is trained to match the target distribution under the squared Hellinger distance \cite{hellinger1909neue},
\begin{equation}
\mathcal{L}_{\text{tag}} =
\tfrac{1}{2} \left\| \sqrt{\hat{\mathbf{p}}_{\text{tag}}} - \sqrt{\mathbf{p}_{\text{tag}}} \right\|_2^2 .
\end{equation}
To further bridge the gap between the textual representation and the collaborative signal, we introduce a stochastic embedding-substitution mechanism at this stage. At each training step, with equal probability, we either feed the \texttt{[CLS]} token embedding of the textual profile into the linear tag head, or replace it with the user's item embedding average obtained from a pretrained and frozen matrix factorization (MF) model similar to \cite{tennenholtz2024demystifying}. 
This stochastic mixing trains a single tag head to be compatible with both representations, which yields two complementary inference modes.
When recommendations are made for a user with a known interaction history, the MF-derived embedding can be used at inference to exploit collaborative signals and achieve higher recommendation accuracy. 
When no interaction history is available, a cold-start user, an edited summary, or a purely text-driven query, the same head decodes directly from the \texttt{[CLS]} embedding of the textual profile, allowing the system to operate in a fully content-grounded regime without relying on any prior interaction. We would like to mention that we have also experimented with using a text LLM to obtain a text-encoding of the user profile. We present the recommendation results from this system in Appendix \ref{sec:llm_encoder}.
\vspace{-5pt}
\paragraph{(3) Collaborative Filtering alignment and scoring.}
CCBR is agnostic to the choice of collaborative filtering backbone; 
we instantiate it with EASE, EDLAE-Full Rank (FR), EDLAE-Low Rank (LR), Linear-DAE, Multi-DAE, Multi-VAE, and MacridVAE, spanning linear item-item, denoising, and variational families. 
Item-item models are factorized by Singular Value Decomposition (SVD) to create an encoder-decoder architecture. 
The Collaborative filtering model that we use as a backbone is trained on user-item interactions in the standard way, and we align the text representation from Stage~2) with the collaborative filtering user latent via an InfoNCE \cite{infonce} objective: 
for each user, $\mathbf{z'}_{\text{text}}$ and $\mathbf{z}_{\text{collab}}$ form the positive pair, and all other users' collaborative latents in the batch serve as negatives. 
The CF decoder is copied and finetuned for the text branch in this stage. Stage-wise optimization settings are summarized in Appendix~\ref{app:training_setup}, and detailed definitions of $\mathcal{L}_{\text{align}}$, $\mathcal{L}_{\text{rec}}$ are provided in Appendix~\ref{app:training_objectives}.

\section{Experiments}
\label{sec:experiments}

\subsection{Experimental setup}
\label{sec:experimentalsetup}
We conduct experiments across three data modalities using a strong generalization approach, in which the model is evaluated on users not seen during training. These include recommendations for image-based clothing products, movie recommendations, and music.
\vspace{-5pt}
\paragraph{Datasets and data modalities.} The datasets that we use are the H\&M fashion dataset \cite{hm}, which contains images for various clothing items, the MovieLens 20M (ML-20M) dataset \cite{ml-20m} (where we use trailers \cite{ml_yt_trailers} for content), and the Million Song Dataset (MSD) \cite{msd} for music recommendations. We follow the strong generalization protocol \cite{mult_dae_vae}, partitioning users disjointly into training, validation, and test sets so that no user in the held-out splits is observed during training. This protocol is a stricter test of generalization than the more common weak generalization setting, in which held-out interactions come from users seen during training. It requires the model to form a useful representation of an entirely new user from their interaction history alone, which aligns directly with CCBR's content-driven design, since a new user's profile is generated from scratch from the items they have interacted with, with no user-specific parameters learned during training. After preprocessing, H\&M contains $1{,}943$ items and $70{,}201$ users (with $1{,}000$ users each held out for validation and test), ML-20M contains $11{,}355$ items and $135{,}269$ users ($5{,}000$ each for validation and test), and MSD contains $31{,}046$ items and $166{,}188$ users ($5{,}000$ each for validation and test).
\vspace{-5pt}
\paragraph{User profile creation pipeline.} To create a user profile for a given user, we first generate a concise natural-language description for each item in their interaction history using a modality-specific MLLM. For the H\&M dataset, item images are described using Qwen3-VL-32B-Instruct \cite{Qwen3-VL}; for ML-20M, movie trailers are described using VideoLLaMA 3 \cite{videollama3}; and for MSD, audio clips are described using MusicFlamingo \cite{ghosh2025music}. This per-item summarization is performed once, offline, and the resulting descriptions are cached and reused across all subsequent training and inference. Once item-level summaries are available, the descriptions of the items in a user's interaction history are aggregated into a single natural-language user profile by an instruction-tuned text-only LLM, for which we use Qwen3-30B-A3B-Instruct-2507 \cite{qwen3technicalreport}. Throughout the pipeline, item identifiers such as titles, product IDs, or artist names are withheld from both the multimodal and text-only models, ensuring that the resulting profiles reflect only content-derived signals and that no pretraining knowledge about specific items leaks into the user representation. 
\vspace{-5pt}
\paragraph{Controllability evaluation pipeline.} To quantify the controllability that CCBR enables, we define a systematic editing pipeline over a fixed set of human-interpretable concepts. The concepts are drawn from the tag vocabulary natively associated with each dataset: movie genres for ML-20M, product categories and colors for H\&M, and tag-based descriptors for  MSD. This choice grounds the evaluation in attributes that are both meaningful to end users and directly verifiable against item-level metadata, so that any shift in the recommendations induced by an edit can be measured against a concrete content axis rather than a free-form notion of similarity. 

Given a user summary, we first identify which of these concepts are explicitly or implicitly mentioned, using an instruction-tuned LLM, Qwen3.5-27B \cite{qwen35}, prompted with the closed concept list and asked to return only concepts that appear in it. We then perform two complementary interventions on the summary, again via the same LLM. In the removal intervention, we take a concept currently present in the profile and rewrite the summary so that all explicit and implicit references to it are excised, while the remainder is preserved; this is repeated for each concept, producing a separate edited summary for each present concept. In the addition intervention, we take a concept that is absent from the profile and rewrite the summary so that a single mention of it is woven in without disturbing the existing references; this too is repeated one concept at a time, producing a separate edited summary for each absent concept. The edited summaries are passed through CCBR, and the resulting recommendations are scored against the catalog's concept matrix: a controllable system should produce monotonic shifts in the recommended items, suppressing items that carry a removed concept and promoting items that carry an introduced one. We defer the precise definitions of the intervention effects, normalization, and aggregation to Section~\ref{sec:controllability}.
Dataset-specific backbone settings are detailed in Appendix~\ref{app:dataset_settings}.

\subsection{Recommendation performance}
\label{sec:performance}

In Table~\ref{tab:results}, we showcase the performance of CCBR across three domains (fashion, music, movies) and seven collaborative filtering backbones. We emphasize that the relevant claim is not uniform performance gains (which would be unrealistic for a deliberately low-bandwidth, human-readable representation) but rather that CCBR remains competitive with strong CF baselines while offering an interpretable and editable user profile, whose controllability we quantify in Section~\ref{sec:controllability}. CCBR rows in Table~\ref{tab:results} further confirm an expected trend: when used in isolation, the text-bottlenecked representation yields lower accuracy than the collaborative backbones, since granular item-level predictions are inherently harder to recover from a compact textual profile than from a collaborative signal trained directly on user-item interactions. Note that, in this and subsequent sections, we only compare with TEARS \cite{tears} with VAE-CF models as the original paper provides results with VAE backbones.

\begin{table*}[ht]
\centering
\caption{Recommendation accuracy on H\&M, MSD, and ML-20M. We evaluate seven collaborative filtering backbones, our proposed CCBR model with a text-bottlenecked user representation aligned via alignment loss. For TEARS models, we consider merged model with $\alpha=0.5$ for fair comparison.}
\label{tab:results}
\vspace{0.5em}
\setlength{\tabcolsep}{4pt}
\renewcommand{\arraystretch}{1.1}
\resizebox{\textwidth}{!}{%
\begin{tabular}{l ccc ccc ccc}
\toprule
& \multicolumn{3}{c}{\textbf{H\&M}} & \multicolumn{3}{c}{\textbf{MSD}} & \multicolumn{3}{c}{\textbf{ML-20M}} \\
\cmidrule(lr){2-4} \cmidrule(lr){5-7} \cmidrule(lr){8-10}
\textbf{Model} & NDCG@100 ($\uparrow$) & R@50  ($\uparrow$)  & R@20 ($\uparrow$) & NDCG@100 ($\uparrow$) & R@50 ($\uparrow$) & R@20 ($\uparrow$) & NDCG@100 ($\uparrow$) & R@50 ($\uparrow$) & R@20 ($\uparrow$) \\
\midrule
EASE           & 0.2621 & 0.2440 & 0.1638 & 0.4519 & 0.3792 & 0.4290 & 0.4909 & 0.5162 & 0.4226 \\
CCBR (Ours)          & 0.2212 & 0.2116 & 0.1359 & 0.3271 & 0.2636 & 0.2979 & 0.4629 & 0.4869 & 0.4008 \\
\midrule
EDLAE (FR)     & 0.2693 & 0.2460 & 0.1662 & 0.4616 & 0.3849 & 0.4351 & 0.5028 & 0.5267 & 0.4323 \\
CCBR (Ours)           & 0.2215 & 0.2148 & 0.1360 & 0.3129 & 0.2521 & 0.2824 & 0.4792 & 0.5057 & 0.4108 \\
\midrule
EDLAE (LR)     & 0.2695 & 0.2447 & 0.1661 & 0.4619 & 0.3858 & 0.4373 & 0.5032 & 0.5275 & 0.4338 \\
CCBR (Ours)           & 0.2215 & 0.2148 & 0.1360 & 0.3233 & 0.2609 & 0.2946 & 0.4802 & 0.5070 & 0.4119 \\
\midrule
Linear-DAE     & 0.2372 & 0.2234 & 0.1432 & 0.3830 & 0.3129 & 0.3499 & 0.4982 & 0.5316 & 0.4261 \\
CCBR (Ours)          & 0.2069 & 0.2013 & 0.1211 & 0.3146 & 0.2544 & 0.2819 & 0.4667 & 0.5011 & 0.3904 \\
\midrule
Multi-DAE      & 0.2160 & 0.2080 & 0.1286 & 0.3632 & 0.3000 & 0.3246 & 0.5042 & 0.5423 & 0.4310 \\
CCBR (Ours)          & 0.2098 & 0.2055 & 0.1264 & 0.3300 & 0.2687 & 0.2988 & 0.4787 & 0.5282 & 0.4055 \\
\midrule
Multi-VAE      & 0.2244 & 0.2135 & 0.1360 & 0.3843 & 0.3144 & 0.3523 & 0.5148 & 0.5468 & 0.4383 \\
TEARS           & 0.2006 & 0.1964 & 0.1161 & 0.2946 & 0.2357 & 0.2602 & 0.4836 & 0.5116 & 0.4105 \\
CCBR (Ours)           & 0.2106 & 0.2064 & 0.1252 & 0.3376 & 0.2746 & 0.3065 & 0.4964 & 0.5370 & 0.4263 \\
\midrule
MacridVAE      & 0.2285 & 0.2206 & 0.1413 & 0.3873 & 0.3151 & 0.3542 & 0.4957 & 0.5218 & 0.4111 \\
TEARS           & 0.1906 & 0.1855 & 0.1081 & 0.2731 & 0.2158 & 0.2358 & 0.4582 & 0.4814 & 0.3872 \\
CCBR   (Ours)         & 0.2016 & 0.1992 & 0.1188 & 0.3247 & 0.2641 & 0.2942 & 0.4829 & 0.5204 & 0.4111 \\
\bottomrule 
\end{tabular}%
}
\end{table*}

\subsection{Controllability Evaluation via Systematic User Summary Editing}
\label{sec:controllability}

Recommendation accuracy alone does not establish that CCBR's text bottleneck is \emph{causally engaged}: a text summary could, in principle, carry information that the CF backbone ignores at inference time. To test whether modifying the natural-language user profile produces a faithful change in the recommended items, we design a pair of counterfactual interventions, Leave-One-Out (LOO) removal and Add-One-In (AOI) addition. A controllable recommendation system should respond monotonically to these edits: removing an attribute from the summary should suppress items carrying it, and adding one should promote them.
\vspace{-5pt}
\paragraph{Setup and notation.}
We instantiate \emph{attributes} from the tag vocabularies natively associated with each dataset. For movies (ML-20M), attributes correspond to movie genres ($C=18$); for H\&M clothing products, we use dominant colors ($C=16$) and clothing categories ($C=51$); for music (MSD), we use tag-based genre descriptors ($C=80$). Let $\mathcal{A} = \{a_1, \dots, a_C\}$ denote this attribute set, and let $\mathcal{A}_i \subseteq \{1, \dots, C\}$ denote the indices of attributes carried by item $i$. For every user $u$, we have a natural-language summary $s_u$ produced by the profile generator pipeline of Section~\ref{sec:experimentalsetup}, and an attribute index set $\mathcal{A}_u \subseteq \{1, \dots, C\}$ obtained by prompting an instruction-tuned LLM to identify which attributes are explicitly or implicitly mentioned in $s_u$.
Given a summary $s$, the trained CCBR model produces a score vector $\mathbf{y}(s) \in \mathbb{R}^N$ over the item catalog. We measure the effect of an edit by comparing the number of top-$K$ recommended items that carry the targeted attribute before and after the edit. For each user $u$, we construct two kinds of edited summaries using the same instruction-tuned LLM.
\vspace{-5pt}
\paragraph{Add-One-In (AOI) additions and Leave-One-Out (LOO) removals.}
An addition edit takes an attribute $c$ that is absent from $\mathcal{A}_u$ and weaves a single mention of $a_c$ into $s_u$ without disturbing the existing references, yielding $\tilde{s}_{u}^{+c}$. 
A removal edit takes an attribute $c$ that is present in $\mathcal{A}_u$ and rewrites $s_u$ so that all explicit and implicit references to $a_c$ are excised while the remainder is preserved, yielding $\tilde{s}_{u}^{\setminus c}$. Each edit targets one attribute at a time, so a user with $n_u = |\mathcal{A}_u|$ mentioned attributes contributes $n_u$ removal edits and $C - n_u$ addition edits. Users with $n_u < 2$ are excluded from the removal analysis, since no attribute remains after the edit. Note that these edits are achieved through prompting an LLM. We assess the locality of H\&M color and category edits using two independent LLM-as-a-judge models, finding that more than $98\%$ of edits preserve non-target profile information (Appendix~\ref{app:edit_locality}).
The exact prompt are shown in Appendix~\ref{app:prompt_details}.
\vspace{-5pt}
\paragraph{Measuring the effect of an edit.}
For a summary $s$, we measure how strongly attribute $c$ is represented in the top-$K$ recommendations by the per-attribute recall
\begin{equation}
    r^{(K)}_c(s) \;=\; \frac{1}{K}\sum_{i \in \mathrm{Top}_K(s)} \mathbf{1}\!\bigl[a_c \in \mathcal{A}_i\bigr],
    \label{eq:recall_attribute}
\end{equation}
i.e., the fraction of top-$K$ items that carry attribute $c$. The per-user effect of an edit is the difference in this recall between the edited and original summaries,
\begin{equation}
    \Delta r^{(K)}_c(u) \;=\; r^{(K)}_c(\tilde{s}_u^{\bullet c}) - r^{(K)}_c(s_u),
\end{equation}
where $\tilde{s}_u^{\bullet c}$ denotes the edited user profile (either $\tilde{s}_u^{+c}$ for additions or $\tilde{s}_u^{\setminus c}$ for removals). A controllable system should yield negative $\Delta r^{(K)}_c$ for removals and positive $\Delta r^{(K)}_c$ for additions. We aggregate these per-user differences by first averaging over all users for whom the attribute was a valid target (present users for removal, absent users for addition), and then averaging across attributes to obtain a global controllability score at cutoff $K$.

\begin{figure}[ht]
    \centering
    \includegraphics[width=0.49\linewidth]{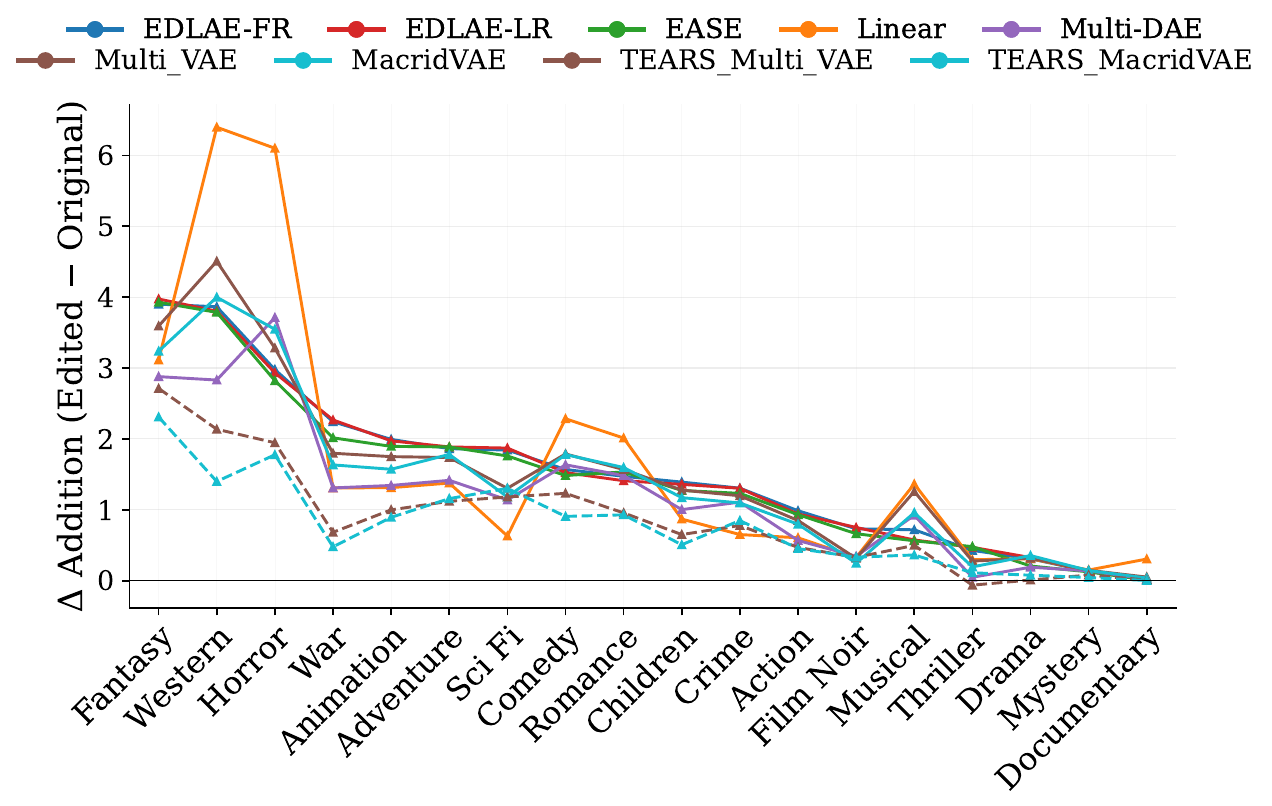}
    \hfill
    \includegraphics[width=0.49\linewidth]{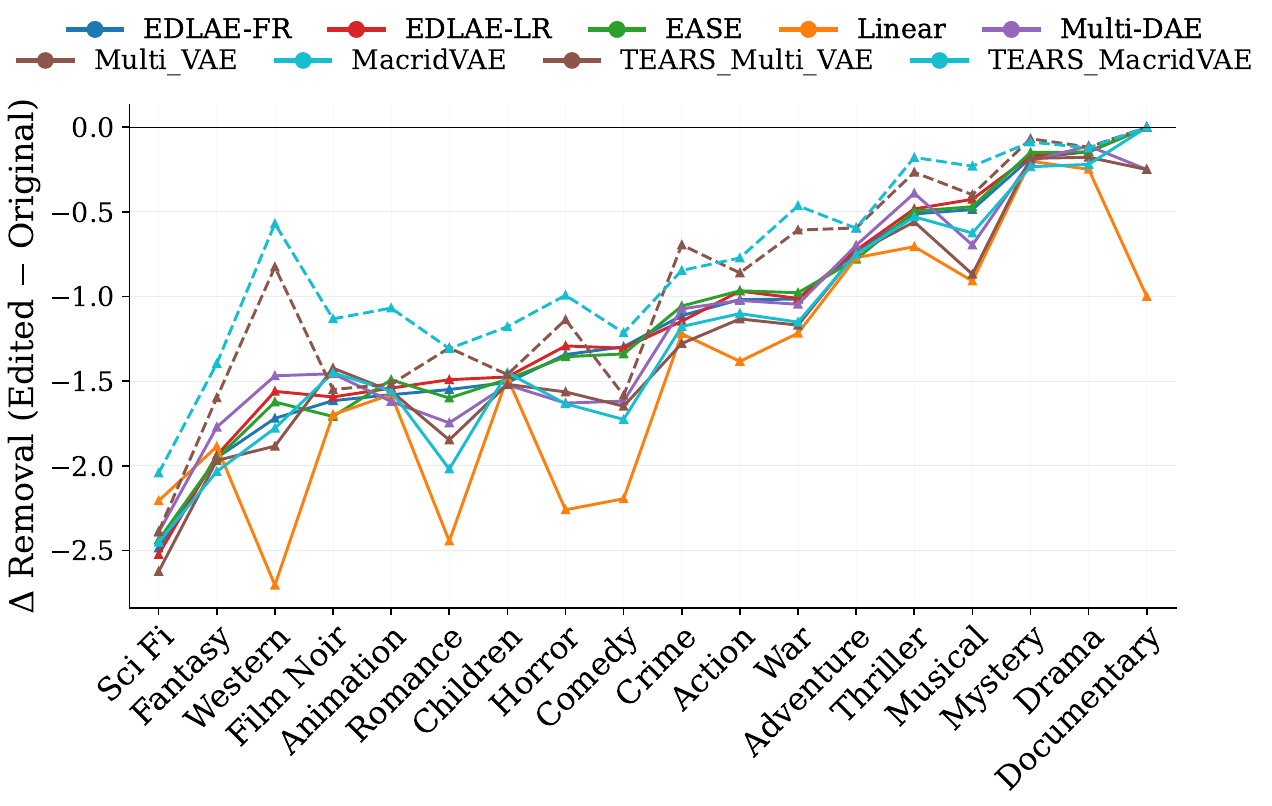}
    \caption{{Controllability (@top-20) on ML-20M across seven collaborative filtering backbones (solid) and two TEARS-trained variants (dashed, same colors as their backbones). Curves show the difference (edited $-$ original) in the per-genre share of recommended items, for addition \textbf{(left)} and removal \textbf{(right)}. Our models consistently outperform their TEARS equivalents: addition deltas are larger (more positive) and removal deltas are larger in magnitude (more negative), meaning summary edits steer the recommender more strongly in the intended direction.}}
    \label{fig:movie-controllability}
\end{figure}
\vspace{-5pt}
\paragraph{Controllability analysis.}
In Figure \ref{fig:movie-controllability}, we showcase the controllability results obtained with CCBR applied to seven different collaborative filtering methodologies presented earlier. We selectively target a concept shown on the x-axis of each panel. We conduct experiments to remove and add concepts. We report the aggregate concepts that are derived from the recommended items after the edit on the y-axis of Figure \ref{fig:movie-controllability}. 

We observe that CCBR is able to very significantly remove the targeted concepts as showcased in the right panel of Figure \ref{fig:movie-controllability}. We also see that, for additions, CCBR manages to add certain movie genres to a significant degree and others to a lesser degree. This difference is potentially due to differences in frequencies across movie genres and the presence of multiple-labels for a given movie. For instance the drama genre ($5{,}723$ occurrences across $11{,}355$ items) is a genre relatively harder to remove or insert because of its high prevalence, and large number of interactions with other genres (e.g., about $1{,}200$ movies with romance tag also have the drama tag). On the other extreme, only few items have `documentary' tag (775), and only four users in the intervention set have watched movies with the documentary tag. We would like to also emphasize that for VAE-based collaborative filtering backbones, CCBR is able to more significantly affect the output compared to TEARS. We also show the editing results for clothing items and music in Appendix \ref{app:hm-plots} and \ref{app:msd-plots}. 

\subsection{Editing the user summary with multimodal inputs}
\label{sec:loi_unseen}

\begin{figure}[ht]
    \centering
    \includegraphics[width=0.85\linewidth]{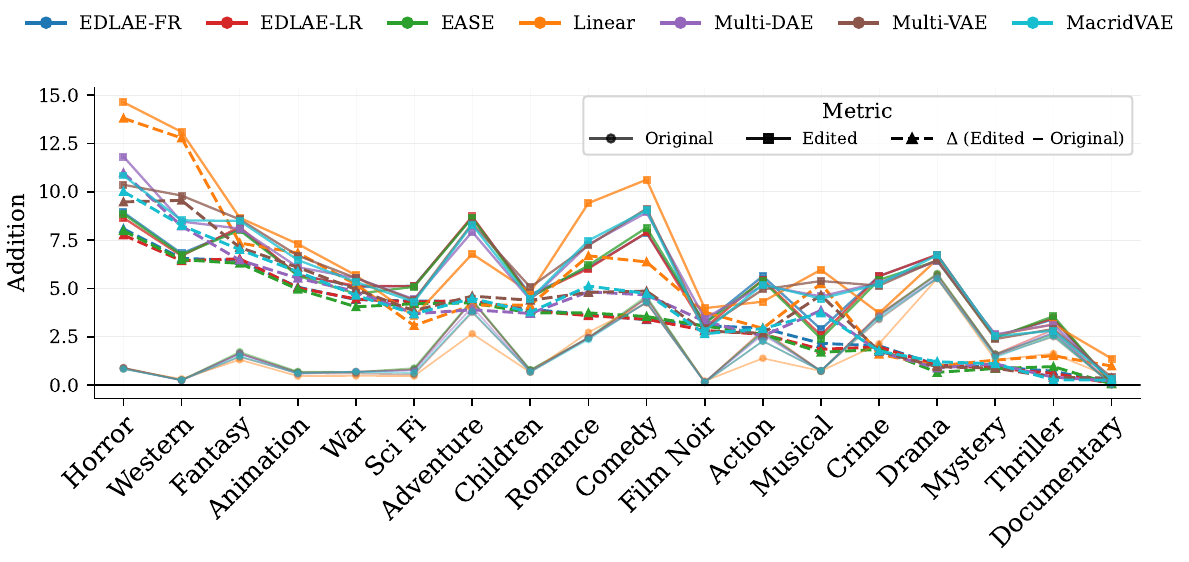}
    \caption{Showcasing controllability on adding unseen items on the ML-20M dataset. We show the controllability for addition, for seven different collaborative filtering backbones as shown in the legend.}
    \label{fig:movie-multimodal}
\end{figure}

The addition intervention of Section~\ref{sec:controllability} edits the user summary at the text level: an instruction-tuned LLM is prompted to insert a single mention of a target attribute into the user summary $s_u$, or to excise references to it for removals. While this is sufficient to test whether the text bottleneck responds to lexical changes, it leaves open the more demanding question of whether CCBR can ingest \emph{new content} through multimodal inputs (images, videos, or audio) that were not seen at training time and route it faithfully through the user profile. We address this with a complementary intervention in which the addition is performed at the item level, using items held out from the catalog during training.

For an additional edit, we maintain a pool of items absent from both the CF training set and the user histories used for profile generation, and obtain their content-derived descriptions using the same modality-specific pipeline as in Section~\ref{sec:method}. For a user $u$ and a target attribute that is not already present in the user's summary ($c \notin \mathcal{A}_u$), we sample unseen items carrying the new concept $a_c$, and re-summarize to produce an edited profile $\tilde{s}_u^{+c}$ that reflects the newly injected items. The key difference from the text-level intervention is that the new concept enters the profile via the modality-specific encoder rather than through direct lexical insertion, allowing us to measure whether CCBR can pick up genuinely unseen content from raw multimodal signals.

This experiment provides two guarantees that the text-level addition cannot. First, it isolates a genuine cold-start regime: the inserted item has no CF representation, so any influence on the recommendation list must be mediated by the text channel. Second, it tests the full pipeline end-to-end (modality encoder, profile editor, text encoder, and CF-aligned decoder) under the same conditions in which CCBR would be deployed when new items enter the catalog after training.
We showcase the obtained editing results in Figure \ref{fig:movie-multimodal}. We observe that with multimodal inputs, CCBR is able to obtain a level of controllability similar to the level obtained with text-only summary edits. Results for the unseen setting on H\&M and MSD are shown in Appendices~\ref{app:hm-plots} and~\ref{app:msd-plots}.

\subsection{Steering the Model with Negative Preferences}
\label{sec:ml20m_negative}
\vspace{-0.5em}
In Section~\ref{sec:controllability} we examined controllability through the removal of concepts from the text summary. That is, we tested whether CCBR can be steered by \emph{negating} a concept, i.e., explicitly stating that the user dislikes the targeted attribute rather than merely deleting references to it. The H\&M and MSD datasets expose only positive interaction signals (purchases, plays), and the user summaries produced in Section~\ref{sec:method} accordingly describe taste in purely affirmative terms. ML-20M, by contrast, provides explicit five-star ratings, giving us access not only to what a user liked but also to what they actively disliked. This asymmetry lets us probe whether CCBR's text bottleneck can carry \emph{signed} preference information and whether the CF backbone responds to it as expected.

\begin{figure}[ht]
    \centering
    \includegraphics[width=0.90\linewidth]{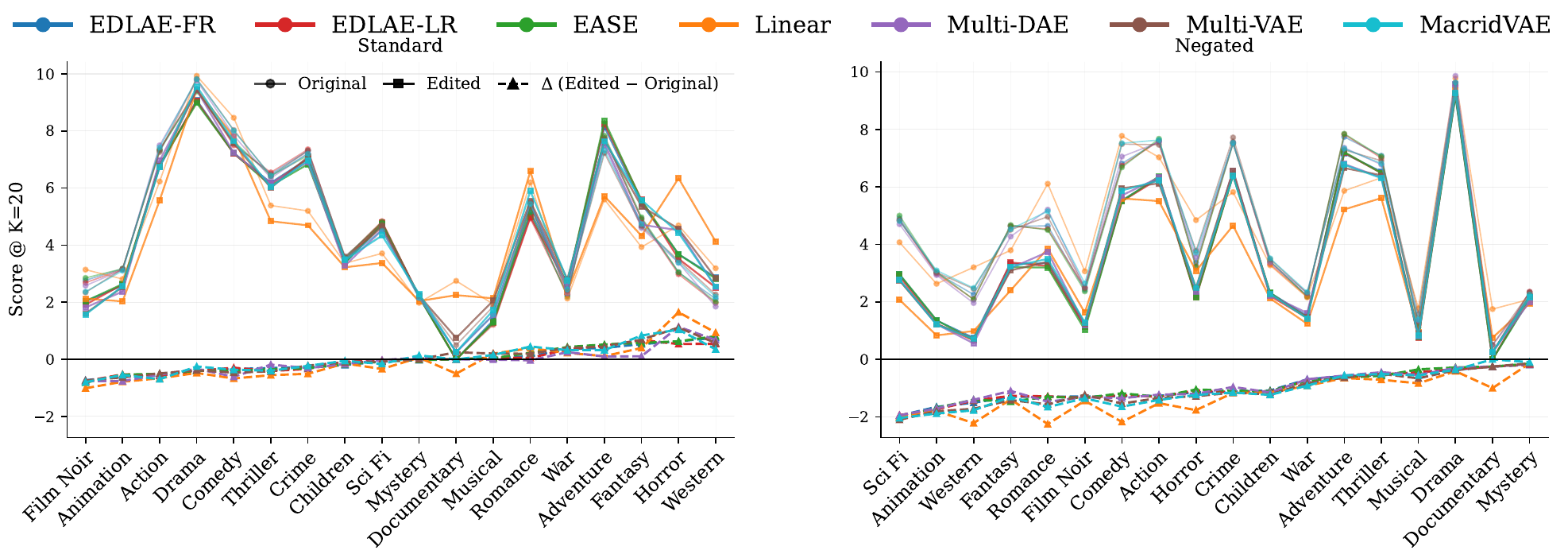}
    \caption{Negation intervention on ML-20M, where the targeted genre in the user summary is rewritten as a dispreference (e.g., ``the user enjoys horror'' $\to$ ``the user dislikes horror''). Solid lines show per-genre recall at $k=20$ for the original and edited summaries; dashed lines show the difference between them. Standard model (\textbf{left}) is trained on positive summaries while negated model (\textbf{right}) is trained on signed summaries.}
    \label{fig:movie-neg}
    \vspace{-0.5em}
\end{figure}
\vspace{-5pt}
\paragraph{Summary generation with negative preferences.} We treat ratings $\geq 4$ as positive interactions and additionally retain ratings $\leq 2$ as \emph{negative} interactions. For each user we partition the history into a positive set $\mathcal{H}_u^{+}$ and a negative set $\mathcal{H}_u^{-}$, and prompt the profile-generation LLM with both sets, instructing it to summarize $\mathcal{H}_u^{+}$ in affirmative language (\emph{``the user likes...''}) and $\mathcal{H}_u^{-}$ in dispreference language (\emph{``the user dislikes...''}). 
As in the positive-only setting, item identifiers are withheld, and the LLM only sees content-derived descriptions, so the resulting profiles remain free of pretraining leakage. CF training is left unchanged: only positively-rated items contribute to $\mathcal{H}_u^{+}$ used by the collaborative backbone, ensuring that any improvement attributable to the signed summary comes through the text channel.

Figure~\ref{fig:movie-neg} compares the negation intervention on two CCBR variants: a \emph{Standard} model trained only on positive interactions, and a \emph{Negated} model trained on signed summaries. Only the Negated model produces a consistent downward shift in the targeted genre's top-$K$ recall, confirming that the text bottleneck transmits signed preference information end-to-end and that the CF backbone responds to negative steering in the expected direction.
\vspace{-5pt}
\paragraph{Computational cost.}
We separate two regimes for computational cost comparison: a one-time offline cost per item, and the online cost of modeling a new user. The multimodal description of each item is computed once and cached, then reused across all training and inference, so it never recurs at serving time; for the video modality, VideoLLaMA 3 takes $4672.9 \pm 2615.4$ ms per item on a single H100-80GB ($\sim$14.7 GPU-hours for the $11{,}355$-item ML-20M catalog), with the image and audio encoders at comparable per-item cost. At serving time, we simulate $100$ users per modality (batch size $1$) with $\mathrm{Uniform}(10, 30)$ interacted items on a single A100-80GB, across three stages: (A) generating the user summary, (B) encoding it, and (C) recommending top-20 items (Table~\ref{tab:cost}). The recurring recommendation-time cost is negligible, embedding takes $\sim$20 ms and scoring under $1$ ms, independent of catalog size; the only substantial costs are the cached per-item description and the per-user summary generation ($\sim$10--13 s), incurred once when a profile is created or edited and not on every recommendation. 

\begin{table}[ht]
\centering
\caption{Online (serving-time) latency per new user across the three CCBR stages, on a single A100-80GB over $100$ simulated users per modality (batch size $1$), each with $\mathrm{Uniform}(10, 30)$ interacted items.}
\label{tab:cost}
\vspace{0.5em}
\setlength{\tabcolsep}{8pt}
\renewcommand{\arraystretch}{1.1}
\begin{tabular}{l ccc}
\toprule
\textbf{Modality (\#items)} & \textbf{A: summary (ms)} & \textbf{B: embed (ms)} & \textbf{C: recommend (ms)} \\
\midrule
Image (1,943 items)   & $10611.8 \pm 1000.6$ & $20.16 \pm 12.55$ & $0.104 \pm 0.010$ \\
Audio (31,046 items)  & $12748.9 \pm 974.6$  & $21.36 \pm 13.87$ & $0.317 \pm 0.960$ \\
Video (11,355 items)  & $10673.2 \pm 880.4$  & $20.07 \pm 14.82$ & $0.251 \pm 0.859$ \\
\bottomrule
\end{tabular}
\end{table}

\section{Conclusions}

We present CCBR, a content-based recommendation framework that introduces controllability to standard collaborative filtering models through an editable, natural-language user profile. By verbalizing items with modality-specific foundation models (images, audio, or video), aggregating them into a textual profile via an instruction-tuned LLM, and aligning the resulting representation with a CF backbone via InfoNCE, CCBR creates a human-readable bottleneck that users can inspect and modify, while preserving the catalog grounding and ranking quality of established collaborative filtering methods.
Across three modalities (clothing images, movie trailers, music audio) and seven CF backbones, 
we observe that CCBR does not significantly deteriorate the recommendation performance while adding the ability to steer the system towards user preferences. 

Our systematic editing experiments show that CCBR is causally engaged: leave-one-out and add-one-in interventions on the user summary produce monotone, attribute-faithful shifts in the recommended items. The framework further supports multimodal interventions by ingesting previously unseen items through the modality-specific encoder and reflecting them in the recommendations. The model also supports dis-preference statements that can suppress the targeted concept. Together, these results indicate that CCBR offers a practical path toward recommendation systems that are accurate, interpretable, and directly steerable by their users.

Taken together, our results point to controllability as a property that can be added to existing recommenders at modest cost rather than as a competing objective. Across three modalities and seven collaborative filtering backbones, the editable text bottleneck preserves most of the ranking quality of the underlying models while exposing a representation that users can read and intervene on, and the leave-one-out, add-one-in, multimodal, and negative-preference interventions all move the recommendations in the intended direction. Because the collaborative model enters CCBR only as a scorer aligned to the text latent, the same mechanism applies uniformly across the linear, denoising, and variational families we consider, suggesting that the interpretability and steerability CCBR provides are largely decoupled from the specific choice of backbone.

A natural avenue for future work is to extend CCBR beyond per-user offline profile generation to interactive settings, where the user iteratively refines the textual summary in dialogue with the system, and to study how the framework composes with personalization signals beyond interaction histories, such as explicit goals or contextual constraints.
\newpage

\section*{Acknowledgments}
We acknowledge the support of the Natural Sciences and Engineering Research Council of Canada (NSERC) and the Digital Research Alliance of Canada (alliancecan.ca). Cem Subakan, and Mirco Ravanelli are respectively supported by the NSERC projects RGPIN-2023-05759, and RGPIN-2022-05088. Mirco Ravanelli also is supported by Translated through the Imminent research grant. Emiliano Penaloza acknowledges the support of the NSERC PGS-D grant and the Bourse en intelligence artificielle provided by Université de Montréal. Laurent Charlin recognizes the support of the Canada CIFAR AI Chair Program, the Canada First Research Excellence Fund and IVADO. We have not received any support for this work except the regular research funding indicated above, and we declare no competing interests.

\bibliographystyle{unsrt}
\bibliography{refs}

\newpage
\appendix
\section{Training Details}
\label{app:training}

\subsection{General Training Setup}
\label{app:training_setup}
For each CF backbone, we perform hyperparameter tuning by sweeping between $30$ and $50$ configurations, with the exact number depending on the model and dataset.

We begin by training a matrix factorization (MF) model on the training interactions and store the resulting $1024$-dimensional item embedding table. A weight decay term is applied during MF training so that the token norm distribution of the learned embeddings matches that of the text encoder. After this step, only the item embedding table is retained.

In the first stage, the text encoder backbone is finetuned with a learning rate of $5\mathrm{e}{-6}$ and the linear heads with a learning rate of $3\mathrm{e}{-4}$ for $100$ epochs, using only the Hellinger objective. We use a batch size of $64$ and, with probability $0.5$, replace the \texttt{[CLS]} embedding with the mean of the user's item embeddings. We also apply dropout with rate $0.2$ to user interactions.

In the second stage, the text encoder backbone learning rate is lowered to $3\mathrm{e}{-6}$. The CF decoder is copied and finetuned for the text branch with a learning rate of $3\mathrm{e}{-5}$, while the loss coefficient for tag prediction is reduced by a factor of $10$. Dropout on user interactions is increased to $0.6$ in this stage. The linear head mapping the text encoder output to the text latent is trained with a learning rate of $3\mathrm{e}{-4}$. We use a batch size of 96 in this stage.

For reconstruction, we use normalized MSE loss for EASE, EDLAE-LR, and EDLAE-FR, and a multinomial objective for the remaining models. For EASE, EDLAE-LR, EDLAE-FR, Linear-DAE, and Multi-DAE, we apply $\ell_2$ normalization to the latents and use InfoNCE alone as the alignment loss. For VAE-based backbones, we omit $\ell_2$ normalization and add a Smooth-L1 loss to match the norms of the text and VAE latents.

\subsection{Training Objectives}
\label{app:training_objectives}
We now formalize the losses used in each training stage. Let $\mathbf{x} \in \{0,1\}^{|\mathcal{I}|}$ denote a user's binary interaction vector over the item set $\mathcal{I}$, and let $\hat{\mathbf{x}}$ denote the reconstruction logits produced by the CF decoder of the text branch. Let $\mathbf{z'}_{\text{text}}$ denote the projected text latent and $\mathbf{z}_{\text{collab}}$ the user latent of the frozen CF backbone, both in $\mathbb{R}^{d}$, where $d$ is the latent dimensionality of the backbone (Appendix~\ref{app:dataset_settings}).

\textbf{Tag loss.} $\mathcal{L}_{\text{tag}}$ is the squared Hellinger distance between the predicted and target tag distributions $\hat{\mathbf{p}}_{\text{tag}}$ and $\mathbf{p}_{\text{tag}}$, as defined in Section~\ref{sec:method}.

\textbf{Alignment loss.} For a batch of $B$ users, the InfoNCE objective with temperature $\tau$ is
\begin{equation}
    \mathcal{L}_{\text{InfoNCE}} \;=\; -\frac{1}{B} \sum_{i=1}^{B} \log \frac{\exp\!\left(\langle \tilde{\mathbf{z}}^{\prime(i)}_{\text{text}}, \tilde{\mathbf{z}}^{(i)}_{\text{collab}} \rangle / \tau\right)}{\sum_{j=1}^{B} \exp\!\left(\langle \tilde{\mathbf{z}}^{\prime(i)}_{\text{text}}, \tilde{\mathbf{z}}^{(j)}_{\text{collab}} \rangle / \tau\right)}.
\end{equation}
For non-VAE backbones (EASE, EDLAE-LR, EDLAE-FR, Linear-DAE, Multi-DAE), both latents are $\ell_2$-normalized, $\tilde{\mathbf{z}} = \mathbf{z}/\|\mathbf{z}\|_2$, and $\mathcal{L}_{\text{align}} = \mathcal{L}_{\text{InfoNCE}}$. For VAE-based backbones (Multi-VAE, MacridVAE), we omit the normalization, $\tilde{\mathbf{z}} = \mathbf{z}$, and additionally match the latent norms with a Smooth-L1 term:
\begin{equation}
    \mathcal{L}_{\text{align}} \;=\; \mathcal{L}_{\text{InfoNCE}} \;+\; \mathrm{SmoothL1}\!\left(\|\mathbf{z'}_{\text{text}}\|_2,\, \|\mathbf{z}_{\text{collab}}\|_2\right).
\end{equation}

\textbf{Reconstruction loss.} For EASE, EDLAE-LR, and EDLAE-FR, we use a normalized MSE loss between the decoder logits and the interaction vector:
\begin{equation}
    \mathcal{L}_{\text{nMSE}} \;=\; \frac{1}{|\mathcal{I}|} \left\| \frac{\hat{\mathbf{x}}}{\|\hat{\mathbf{x}}\|_2} - \frac{\mathbf{x}}{\|\mathbf{x}\|_2} \right\|_{2}^{2}.
\end{equation}
For the remaining backbones (Linear-DAE, Multi-DAE, Multi-VAE, and MacridVAE), we use the multinomial log-likelihood:
\begin{equation}
    \mathcal{L}_{\text{mult}} \;=\; -\sum_{k=1}^{|\mathcal{I}|} x_{k} \, \log \mathrm{softmax}(\hat{\mathbf{x}})_{k}.
\end{equation}

\textbf{Stage 1.} The text encoder and linear heads are trained with the tag loss alone, $\mathcal{L}_{\text{stage 1}} = \mathcal{L}_{\text{tag}}$.

\textbf{Stage 2.} The tag loss is down-weighted by a factor of $10$, and the alignment and reconstruction losses are added with weights $\alpha_{2}$ and $\alpha_{3}$:
\begin{equation}
    \mathcal{L}_{\text{stage 2}} \;=\; 0.1 \cdot \mathcal{L}_{\text{tag}} \;+\; \alpha_{2} \cdot \mathcal{L}_{\text{align}} \;+\; \alpha_{3} \cdot \mathcal{L}_{\text{rec}},
\end{equation}
where $\mathcal{L}_{\text{rec}}$ is $\mathcal{L}_{\text{nMSE}}$ for EASE, EDLAE-LR, and EDLAE-FR and $\mathcal{L}_{\text{mult}}$ otherwise.

\subsection{Dataset-specific Settings}
\label{app:dataset_settings}
\textbf{ML-20M.}
We follow the general recipe described above. For the variant that incorporates negated summaries, we additionally train a second MF model on the negative interactions extracted from the user histories. The EASE and EDLAE-FR models are factorized to a $1024$-dimensional latent space via SVD.

\textbf{MSD.}
It contains approximately $31{,}046$ items, and a $1024$-dimensional latent space is insufficient for the linear models in this regime. We therefore use a $4096$-dimensional latent space for EASE, EDLAE-FR, EDLAE-LR, and Linear-DAE on this dataset. For Multi-DAE, Multi-VAE, and MacridVAE, increasing the latent dimensionality did not yield further gains, so we retain the $1024$-dimensional latent space for these backbones.

\textbf{H\&M.}
Roughly one third of the items in H\&M are black, which introduces a substantial class imbalance during the first stage of training. To mitigate this, we augment the training set by intervening on the user summaries to modify or remove explicit references to the color black, and use this augmented set for the first stage of training. The second stage is then trained on the original, unaugmented summaries.

\section{Controllability Details}
\label{app:controllability_details}

\subsection{Independent Validation of Edit Locality}
\label{app:edit_locality}

We assess whether H\&M color and category edits preserve non-target profile information using two independent LLM-as-a-judge models, GLM-4 32B-0414 \cite{glm432b0414} and Gemma-4 12B \cite{gemma4}. For each (before, after, target attribute) triple, each judge returns a binary \texttt{preserved} label. An edit is judged preserved only when all non-target preferences retain their meaning and no new preferences are introduced. As shown in Table~\ref{tab:edit_locality_table}, both judges consistently find that the edits preserve non-target profile information across addition and removal interventions, with preservation rates above $98\%$ for every intervention. This suggests that the edits are generally localized to the target attribute rather than substantially rewriting the user profile.

\begin{table}[ht]
\centering
\caption{Independent LLM-as-a-judge evaluation of whether H\&M profile edits preserve non-target information. Values report the percentage of edits judged to preserve the remainder of the profile.}
\label{tab:edit_locality_table}
\vspace{0.5em}
\setlength{\tabcolsep}{8pt}
\renewcommand{\arraystretch}{1.1}
\begin{tabular}{lcc}
\toprule
\textbf{Intervention} & \textbf{GLM-4 32B-0414} & \textbf{Gemma-4 12B} \\
\midrule
Remove color    & 100.00\% & 99.83\% \\
Remove category & 99.98\%  & 98.01\% \\
Add color       & 100.00\% & 99.64\% \\
Add category    & 99.99\%  & 99.73\% \\
\bottomrule
\end{tabular}
\end{table}

\newpage
\subsection{Prompt Details}
\label{app:prompt_details}

\textbf{Attribute extraction prompt.}
\begin{tcolorbox}[colback=gray!5, colframe=gray!50, boxrule=0.4pt, arc=1pt, left=4pt, right=4pt, top=3pt, bottom=3pt]
\small\ttfamily\raggedright

"You are a precise movie genre extractor. Given a user watching summary, identify which movie genres are mentioned or clearly implied.\\
IMPORTANT RULES:\\
- Only extract genres that are explicitly mentioned or very strongly implied by the summary.\\
- Map related terms to the closest valid genre using these rules:\\
\hspace*{1em}- "science fiction" or "sci fi" $\rightarrow$ Sci-Fi\\
\hspace*{1em}- "noir" or "film noir" $\rightarrow$ Film-Noir\\
\hspace*{1em}- "kids" or "family" $\rightarrow$ Children\\
\hspace*{1em}- "animated" or "cartoon" $\rightarrow$ Animation\\
\hspace*{1em}- "suspense" or "psychological thriller" $\rightarrow$ Thriller\\
\hspace*{1em}- "romantic" or "love story" or "rom-com" $\rightarrow$ Romance\\
\hspace*{1em}- "dark comedy" or "satire" or "slapstick" $\rightarrow$ Comedy\\
\hspace*{1em}- "war film" or "military" or "combat" $\rightarrow$ War\\
\hspace*{1em}- "detective" or "whodunit" or "murder mystery" $\rightarrow$ Mystery\\
\hspace*{1em}- "gangster" or "heist" or "mob" or "mafia" $\rightarrow$ Crime\\
\hspace*{1em}- "superhero" or "martial arts" or "kung fu" $\rightarrow$ Action\\
\hspace*{1em}- "spy" or "espionage" $\rightarrow$ Thriller\\
\hspace*{1em}- "zombie" or "slasher" or "supernatural" or "gothic" $\rightarrow$ Horror\\
\hspace*{1em}- "fairy tale" or "mythological" or "magical" $\rightarrow$ Fantasy\\
\hspace*{1em}- "space opera" or "dystopian" or "cyberpunk" or "time travel" $\rightarrow$ Sci-Fi\\
\hspace*{1em}- "cowboy" or "frontier" or "wild west" $\rightarrow$ Western\\
\hspace*{1em}- "biographical" or "biopic" or "courtroom" or "coming of age" $\rightarrow$ Drama\\
\hspace*{1em}- "song and dance" or "concert" $\rightarrow$ Musical\\
\hspace*{1em}- "nature documentary" or "docudrama" $\rightarrow$ Documentary\\
\hspace*{1em}- "quest" or "expedition" or "survival" $\rightarrow$ Adventure\\
- Do NOT hallucinate genres not present or implied in the summary.\\
VALID GENRES:\\
\{genres\}\\
Respond ONLY with valid JSON, no other text:\\
\{"genres": ["Genre1", "Genre2", ...]\}"
\end{tcolorbox}

\textbf{Removal edit prompt.}
\begin{tcolorbox}[colback=gray!5, colframe=gray!50, boxrule=0.4pt, arc=1pt, left=4pt, right=4pt, top=3pt, bottom=3pt]
\small\ttfamily\raggedright
"You are a precise text editor. Your task is to edit a movie watching summary to REMOVE all mentions and implications of ONE specific movie genre, while keeping the rest of the summary natural and coherent.\\
GENRE TO REMOVE: \{genre\_to\_remove\}\\
ORIGINAL SUMMARY:\\
\{summary\}\\
RULES:\\
- Remove all explicit and implicit references to the specified genre.\\
- Do NOT add any new movie genres or attributes.\\
- Keep the overall structure, tone, and remaining content intact.\\
- The edited summary should read naturally, as if that genre was never part of the user's taste.\\
- If removing the genre leaves an awkward gap or broken sentence, smooth it out minimally.\\
- Do NOT remove any other genres or attributes beyond the one specified.\\
Respond with ONLY the edited summary text, nothing else."
\end{tcolorbox}

\textbf{Addition edit prompt.}
\begin{tcolorbox}[colback=gray!5, colframe=gray!50, boxrule=0.4pt, arc=1pt, left=4pt, right=4pt, top=3pt, bottom=3pt]
\small\ttfamily\raggedright
"You are a precise text editor. Your task is to edit a movie watching summary to naturally INCORPORATE mentions of ONE specific new movie genre, while keeping the rest of the summary coherent.\\
GENRE TO ADD: \{genre\_to\_add\}\\
ORIGINAL SUMMARY:\\
\{summary\}\\
RULES:\\
- Weave the specified genre naturally into the existing summary.\\
- It should feel like a genuine part of the user's watching preferences.\\
- Do NOT remove any existing genres or attributes.\\
- Keep the overall structure and tone consistent.\\
- Add minimal text, just enough to incorporate the new genre naturally.\\
- The genre should be clearly identifiable in the edited text (explicitly mentioned or very strongly implied).\\
Respond with ONLY the edited summary text, nothing else."
\end{tcolorbox}

\section{Plots}
\vspace{1.0em}
\subsection{H\&M}
\label{app:hm-plots}

\begin{figure}[ht]
    \centering
    \includegraphics[width=0.90\linewidth]{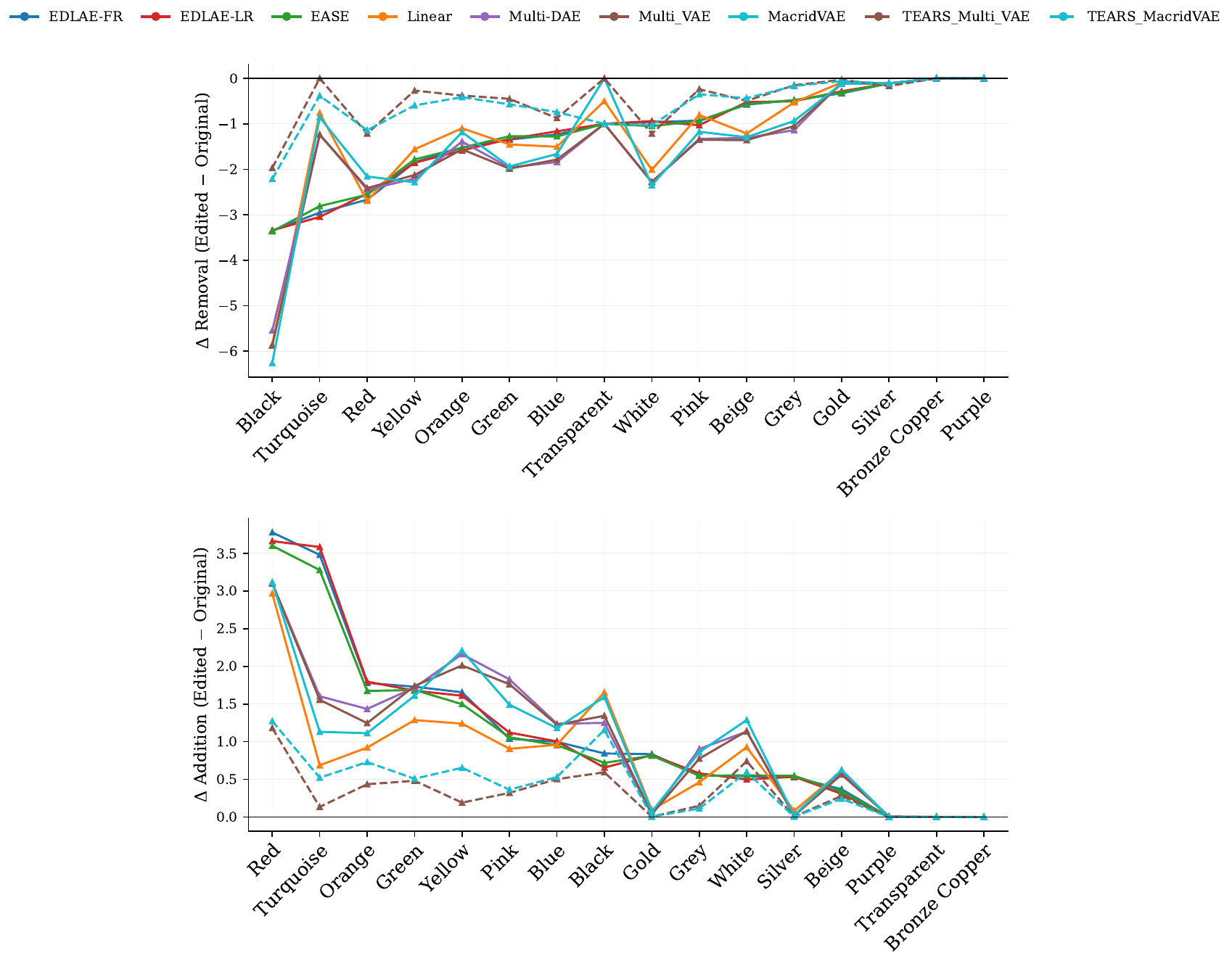}
    \caption{Per-color edit effects on the H\&M dataset at $k{=}20$. Top: $\Delta$ Removal (Edited $-$ Original); bottom: $\Delta$ Addition (Edited $-$ Original). Colors are sorted by frequency. TEARS variants (dashed) respond more weakly than the standard models, especially on rare colors, indicating that gradient-based steering through the text encoder produces smaller but more selective shifts than direct interaction-level edits.}
    \label{fig:hm-color}
\end{figure}

\begin{figure}[ht]
    \centering
    \includegraphics[width=0.90\linewidth]{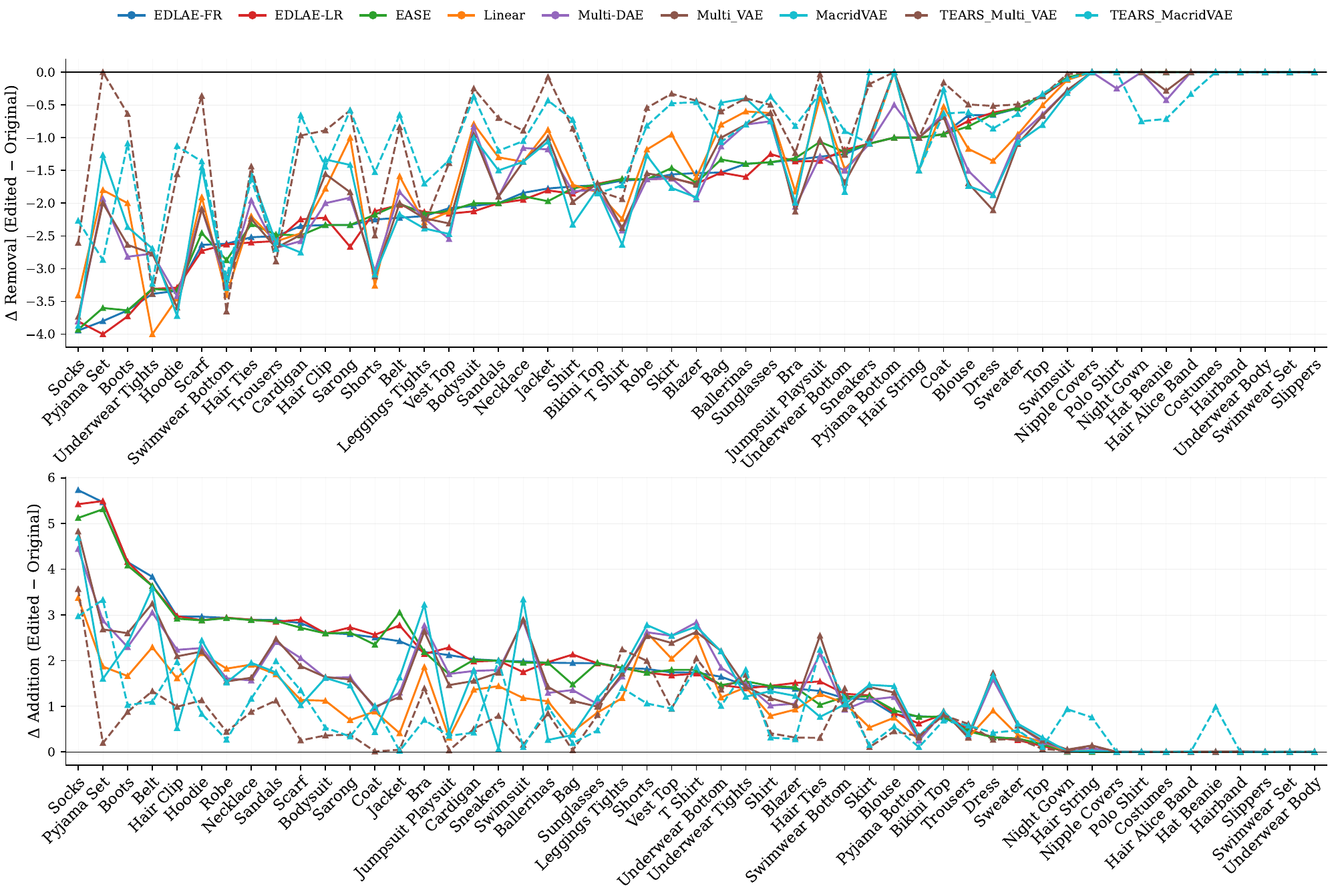}
    \caption{Per-product-type edit effects on the H\&M dataset at $k{=}20$. Top: $\Delta$ Removal; bottom: $\Delta$ Addition. Product types are sorted by frequency. All models respond strongly on frequent items (Socks, Pyjama Set, Boots) and converge to near-zero deltas on rare ones (Slippers, Underwear Body, Swimwear Set), reflecting the sparsity floor below which edits cannot meaningfully influence rankings.}
    \label{fig:hm-product}
\end{figure}

\begin{figure}[ht]
    \centering
    \includegraphics[width=0.97\linewidth]{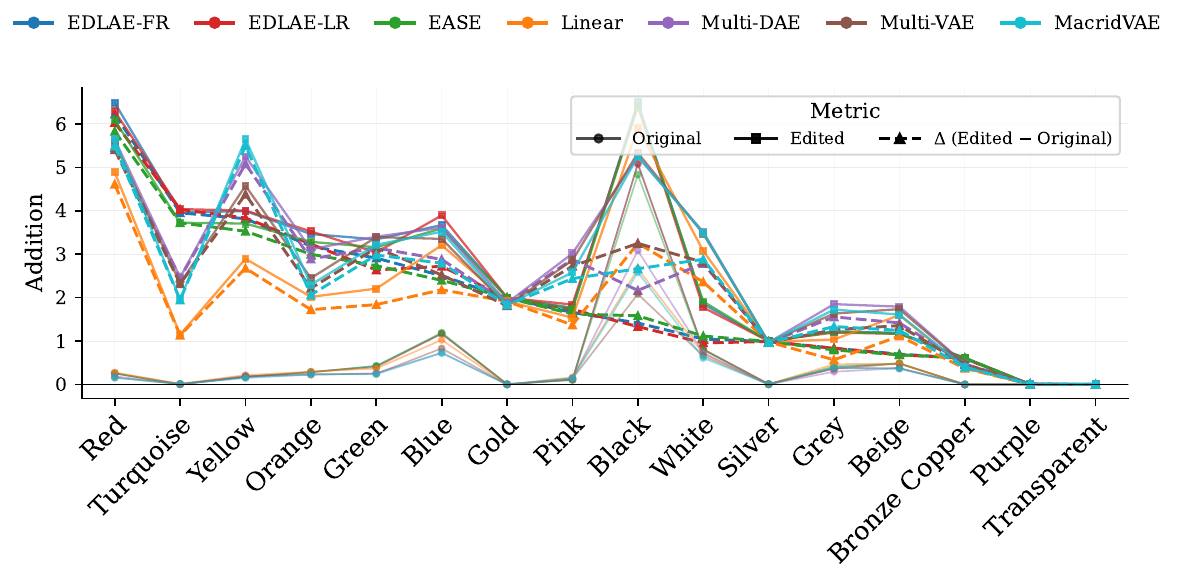}
    \caption{Per-color addition effects on the H\&M dataset at $k{=}20$ for the unseen (multimodal-input) setting. Solid lines show Original and Edited recommendations; dashed lines show $\Delta$ (Edited $-$ Original). Edits propagate cleanly even for colors absent from the user's history, showing that text-conditioned models can introduce previously unseen attributes rather than merely amplify existing ones.}
    \label{fig:hm-color-unseen}
\end{figure}

\begin{figure}[ht]
    \centering
    \includegraphics[width=0.97\linewidth]{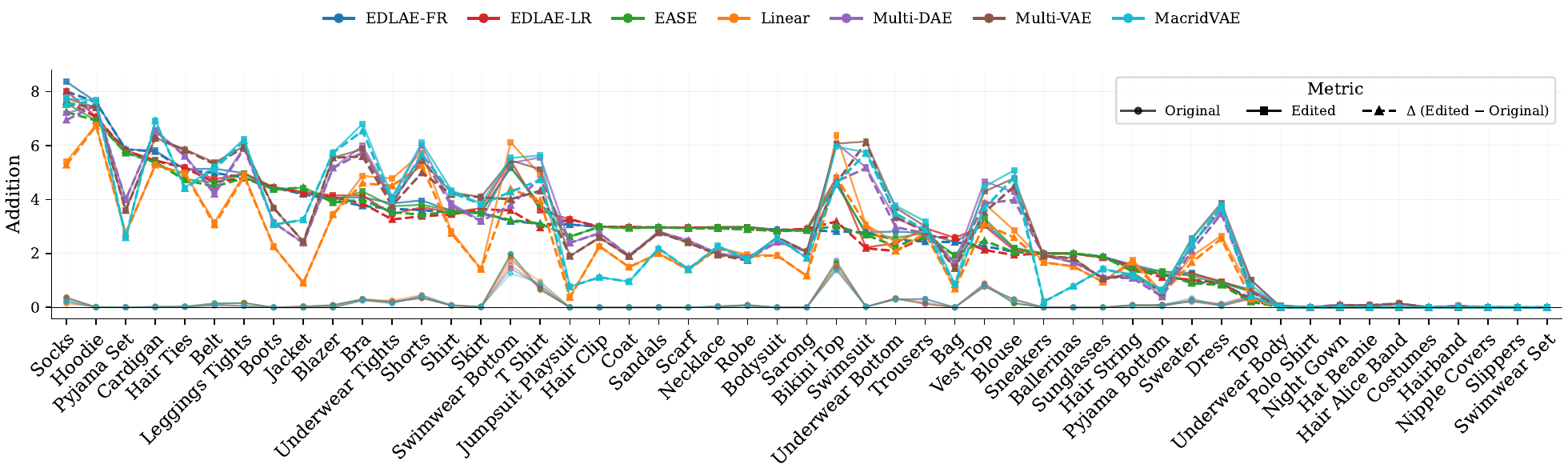}
    \caption{Per-product-type addition effects on the H\&M dataset at $k{=}20$ for the unseen setting. Solid lines: Original and Edited recommendations; dashed: $\Delta$. The largest deltas concentrate on common product types and decay sharply for rare ones, confirming that the unseen-attribute setting inherits the same long-tail behavior observed in the seen case.}
    \label{fig:hm-product-unseen}
\end{figure}

\clearpage
\newpage
\subsection{MSD}
\label{app:msd-plots}

\begin{figure}[ht]
    \centering
    \includegraphics[width=0.97\linewidth]{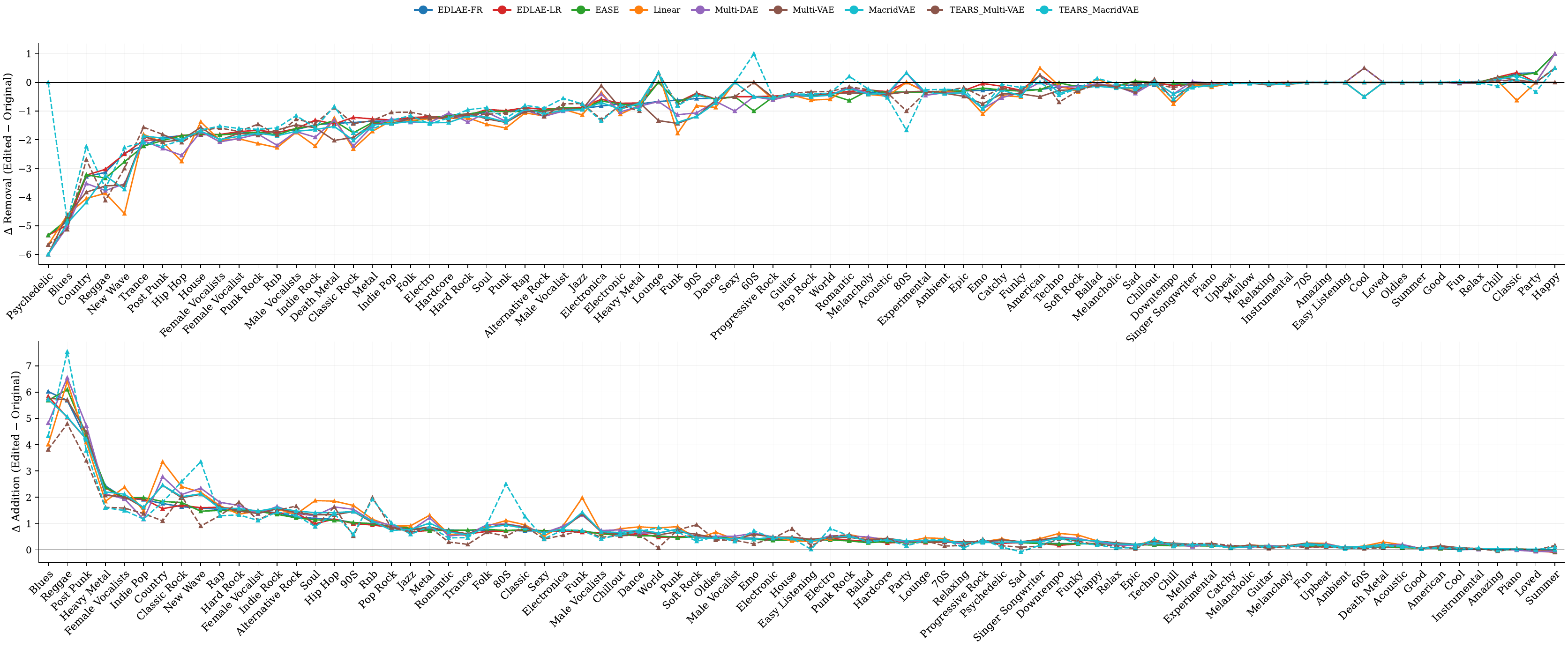}
    \caption{Per-tag edit effects on the MSD (Million Song Dataset) at $k{=}20$, covering 80 music tags spanning genres, eras and mood descriptors. Top: $\Delta$ Removal; bottom: $\Delta$ Addition (Edited $-$ Original); TEARS variants are dashed. Removal effects are largest on Psychedelic, Blues and Country, while addition effects peak on Blues, Reggae and Post Punk. The asymmetry between top removal and top addition tags shows that the two operations engage different parts of the tag space rather than being symmetric inverses.}
    \label{fig:music}
\end{figure}

\begin{figure}[ht]
    \centering
    \includegraphics[width=0.97\linewidth]{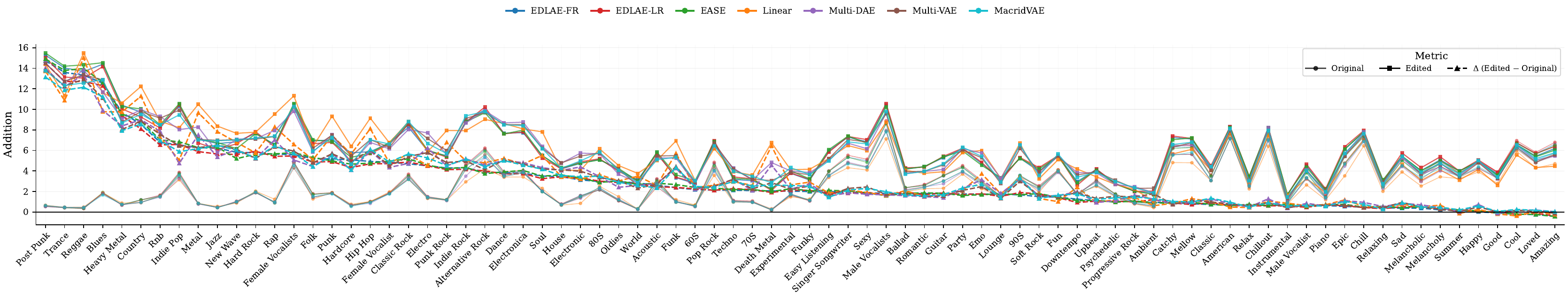}
\caption{Per-tag addition effects on the MSD at $k{=}20$ for the unseen (multimodal-input) setting, across the seven collaborative filtering backbones. Line styles distinguish the Original and Edited recommendations and their difference $\Delta$ (Edited $-$ Original). Injecting unseen items that carry a target tag increases the share of that tag in the recommendations for most tags, with effect sizes varying across tags.}
    \label{fig:music-unseen}
\end{figure}

\section{CCBR with an LLM Text Encoder}
\label{sec:llm_encoder}

In the main experiments, CCBR encodes the natural-language user profile with a BERT-style text encoder (BGE-M3). Here we study an alternative instantiation in which the text encoder is replaced by a similarly-sized LLM encoder, Qwen3.5-0.8B \cite{qwen35}, while the rest of the pipeline is left unchanged and the user profile remains fully editable. Unlike the two-stage recipe of Section~\ref{sec:method}, the LLM encoder is trained in a single stage, where tag supervision, alignment, and reconstruction are optimized jointly rather than split across a tag-pretraining stage followed by a CF-alignment stage.

Table~\ref{tab:llm_encoder} reports the recommendation accuracy obtained with this LLM encoder across all seven collaborative filtering backbones. Relative to the BGE-M3 encoder, the LLM encoder improves NDCG@100 by $+7.7\%$ on average on H\&M and $+14.1\%$ on MSD, while remaining on par on ML-20M ($+1.2\%$). The improvements are most pronounced on MSD, the dataset with the largest catalog ($\sim$31K items), where a stronger text encoder helps the compact textual profile recover more of the granular item-level signal. These results indicate that CCBR is agnostic to the choice of text encoder and benefits directly from a more capable one, without any change to the editable user profile or the single-stage training setup described above.

\begin{table*}[ht]
\centering
\caption{Recommendation accuracy on H\&M, MSD, and ML-20M with the single-stage LLM text encoder (Qwen3.5-0.8B). For each collaborative filtering backbone we report the backbone and the corresponding CCBR model.}
\label{tab:llm_encoder}
\vspace{0.5em}
\setlength{\tabcolsep}{4pt}
\renewcommand{\arraystretch}{1.1}
\resizebox{\textwidth}{!}{%
\begin{tabular}{l ccc ccc ccc}
\toprule
& \multicolumn{3}{c}{\textbf{H\&M}} & \multicolumn{3}{c}{\textbf{MSD}} & \multicolumn{3}{c}{\textbf{ML-20M}} \\
\cmidrule(lr){2-4} \cmidrule(lr){5-7} \cmidrule(lr){8-10}
\textbf{Model} & NDCG@100 ($\uparrow$) & R@50 ($\uparrow$) & R@20 ($\uparrow$) & NDCG@100 ($\uparrow$) & R@50 ($\uparrow$) & R@20 ($\uparrow$) & NDCG@100 ($\uparrow$) & R@50 ($\uparrow$) & R@20 ($\uparrow$) \\
\midrule
EASE           & 0.2621 & 0.2440 & 0.1638 & 0.4519 & 0.3792 & 0.4290 & 0.4909 & 0.5162 & 0.4226 \\
CCBR (Ours)    & 0.2434 & 0.2252 & 0.1500 & 0.3778 & 0.3091 & 0.3490 & 0.4658 & 0.4854 & 0.4034 \\
\midrule
EDLAE (FR)     & 0.2693 & 0.2460 & 0.1662 & 0.4616 & 0.3849 & 0.4351 & 0.5028 & 0.5267 & 0.4323 \\
CCBR (Ours)    & 0.2500 & 0.2303 & 0.1544 & 0.3793 & 0.3103 & 0.3496 & 0.4852 & 0.5092 & 0.4203 \\
\midrule
EDLAE (LR)     & 0.2695 & 0.2447 & 0.1661 & 0.4619 & 0.3858 & 0.4373 & 0.5032 & 0.5275 & 0.4338 \\
CCBR (Ours)    & 0.2466 & 0.2243 & 0.1536 & 0.3830 & 0.3141 & 0.3546 & 0.4717 & 0.4951 & 0.4060 \\
\midrule
Linear-DAE     & 0.2372 & 0.2234 & 0.1432 & 0.3830 & 0.3129 & 0.3499 & 0.4982 & 0.5316 & 0.4261 \\
CCBR (Ours)    & 0.2155 & 0.2020 & 0.1309 & 0.3525 & 0.2889 & 0.3179 & 0.4883 & 0.5226 & 0.4129 \\
\midrule
Multi-DAE      & 0.2160 & 0.2080 & 0.1286 & 0.3632 & 0.3000 & 0.3246 & 0.5042 & 0.5423 & 0.4310 \\
CCBR (Ours)    & 0.2183 & 0.2059 & 0.1316 & 0.3579 & 0.2939 & 0.3154 & 0.5010 & 0.5387 & 0.4273 \\
\midrule
Multi-VAE      & 0.2244 & 0.2135 & 0.1360 & 0.3843 & 0.3144 & 0.3523 & 0.5148 & 0.5468 & 0.4383 \\
CCBR (Ours)    & 0.2110 & 0.2003 & 0.1265 & 0.3696 & 0.3051 & 0.3314 & 0.4847 & 0.5191 & 0.4101 \\
\midrule
MacridVAE      & 0.2285 & 0.2206 & 0.1413 & 0.3873 & 0.3151 & 0.3542 & 0.4957 & 0.5218 & 0.4111 \\
CCBR (Ours)    & 0.2237 & 0.2181 & 0.1370 & 0.3675 & 0.3014 & 0.3291 & 0.4899 & 0.5243 & 0.4143 \\
\bottomrule
\end{tabular}%
}
\end{table*}

\section{Beyond Linear and Autoencoder Backbones}
\label{app:beyond_cf}

CCBR is agnostic to the choice of collaborative backbone: the CF model acts as a frozen scorer aligned to the text latent via InfoNCE, so the controllability layer only requires the backbone to expose a user-level representation that the text latent can be aligned to, after which the decoder is reused to score items. To demonstrate that this design extends beyond the linear item-item and autoencoder families considered in the main text, we integrate two additional backbones that learn fundamentally different user-item interactions: NeuMF \cite{he2017neural}, which learns a nonlinear interaction rather than a simple factorization, and LightGCN \cite{lightgcn}, a graph-convolutional recommender. Both slot into CCBR with no change to the framework; a new backbone only needs to be trained under strong generalization and re-run through the alignment stage.

Table~\ref{tab:beyond_cf} reports results for these backbones and their CCBR counterparts on the H\&M (image) and ML-20M (movie) domains. Beyond preserving recommendation accuracy, we observe that attaching CCBR can in fact improve the final recommendation performance in several cases.

\begin{table}[ht]
\centering
\caption{CCBR applied to non-linear and graph-based backbones (NeuMF and LightGCN) on the H\&M (image) and ML-20M (movie) domains. Both backbones integrate into CCBR without any change to the framework.}
\label{tab:beyond_cf}
\vspace{0.5em}
\setlength{\tabcolsep}{6pt}
\renewcommand{\arraystretch}{1.1}
\resizebox{\textwidth}{!}{%
\begin{tabular}{l ccc ccc}
\toprule
& \multicolumn{3}{c}{\textbf{H\&M}} & \multicolumn{3}{c}{\textbf{ML-20M}} \\
\cmidrule(lr){2-4} \cmidrule(lr){5-7}
\textbf{Model} & NDCG@100 ($\uparrow$) & R@50 ($\uparrow$) & R@20 ($\uparrow$) & NDCG@100 ($\uparrow$) & R@50 ($\uparrow$) & R@20 ($\uparrow$) \\
\midrule
NeuMF             & 0.1842 & 0.1759 & 0.1148 & 0.4485 & 0.4841 & 0.3981 \\
CCBR (Ours)       & 0.1958 & 0.1924 & 0.1175 & 0.4634 & 0.4864 & 0.3981 \\
\midrule
LightGCN          & 0.2041 & 0.1954 & 0.1231 & 0.4187 & 0.4468 & 0.3483 \\
CCBR (Ours)       & 0.2007 & 0.1974 & 0.1221 & 0.4621 & 0.4989 & 0.3958 \\
\bottomrule
\end{tabular}%
}
\end{table}

More broadly, the core mechanism of CCBR is not specific to the CF/MF setting. The controllability layer only requires the backbone to expose a user-level representation that the text latent can be aligned to; once this target exists, the decoder is reused to score items. Two-tower models are an especially natural fit, as the user tower already emits a user embedding that can serve directly as the alignment target, with the item tower playing the role of the decoder. Sequential and transformer-based recommenders can be accommodated similarly, by aligning the text latent to the final user hidden state. The main requirement is simply that a user-side representation exists to align against; architectures that output only an opaque score, with no explicit user vector, would first need such a target to be defined.

\section{Broader Impacts}
\label{app:broader_impacts}

\paragraph{Positive impacts.} CCBR represents each user through a natural-language profile that the user can read and edit, rather than through an opaque latent vector. This makes it possible for users to see how a recommendation system represents them, to correct inaccurate or outdated inferences, and to steer recommendations toward their current goals, for example by temporarily emphasizing a style, genre, or mood that differs from their long-term history. Such transparency and control can increase user agency and trust, and can make it easier for users and auditors to identify when a system has formed an undesirable picture of a user. Because CCBR can be attached to existing collaborative filtering models, these benefits do not require replacing deployed recommenders.

\paragraph{Negative impacts and mitigations.} 

CCBR is an interpretable-by-design recommender. While we see no immediate negative societal impacts, the explicit user profiles it exposes could be misused: they may reveal sensitive user preferences. In addition, LLM-generated profiles can inherit biases from the underlying model. Treating profiles as personal data, auditing them for bias, and letting users view and edit their own profiles can reduce these risks.


\clearpage
\newpage
\section*{NeurIPS Paper Checklist}

\begin{enumerate}

\item {\bf Claims}
    \item[] Question: Do the main claims made in the abstract and introduction accurately reflect the paper's contributions and scope?
    \item[] Answer: \answerYes{}
    \item[] Justification: The abstract and Section~\ref{sec:intro} make three claims: (i) CCBR introduces controllability into a given CF backbone through an editable, content-derived text bottleneck; (ii) the resulting models remain competitive with their underlying CF backbones and outperform the controllable baseline TEARS; and (iii) recommendations can be steered through text edits, previously unseen multimodal inputs, and negative preference statements. Claim (i) is described in Section~\ref{sec:method}. Claim (ii) is supported by Table~\ref{tab:results}, which also reports the accuracy gap to the backbones, most notably on MSD. Claim (iii) is supported by Sections~\ref{sec:controllability}, \ref{sec:loi_unseen}, and~\ref{sec:ml20m_negative}, with additional results in Appendices~\ref{app:hm-plots} and~\ref{app:msd-plots}.

\item {\bf Limitations}
    \item[] Question: Does the paper discuss the limitations of the work performed by the authors?
    \item[] Answer: \answerYes{}
    \item[] Justification: Section~\ref{sec:performance} acknowledges that the text-bottlenecked representation yields lower accuracy than the underlying CF backbones, with the largest gap on MSD, the dataset with the largest catalog. The computational cost analysis (Table~\ref{tab:cost}) reports that generating or editing a user summary takes about 10 to 13 seconds per user; this is a one-time cost per profile, but it remains the dominant serving-time expense. Section~\ref{sec:controllability} notes that controllability is weaker for highly prevalent or rare attributes and that the evaluated edits are produced by prompting an LLM. The Conclusions note that CCBR is currently evaluated in an offline profile-generation setting and leave interactive refinement to future work.

\item {\bf Theory assumptions and proofs}
    \item[] Question: For each theoretical result, does the paper provide the full set of assumptions and a complete (and correct) proof?
    \item[] Answer: \answerNA{}
    \item[] Justification: The paper does not include theoretical results; the contribution is methodological and empirical.

\item {\bf Experimental result reproducibility}
    \item[] Question: Does the paper fully disclose all the information needed to reproduce the main experimental results of the paper to the extent that it affects the main claims and/or conclusions of the paper (regardless of whether the code and data are provided or not)?
    \item[] Answer: \answerYes{}
    \item[] Justification: Section~\ref{sec:method} describes the full CCBR pipeline (profile generation, tag-supervised text encoding, and InfoNCE alignment with the CF backbone). Section~\ref{sec:experimentalsetup} specifies the datasets, the strong generalization protocol with the resulting split sizes, and the exact models used at each stage (Qwen3-VL-32B-Instruct, VideoLLaMA~3, MusicFlamingo, Qwen3-30B-A3B-Instruct-2507, BGE-M3, and Qwen3.5-27B). Appendix~\ref{app:training} reports the stage-wise training setup, learning rates, batch sizes, loss definitions, and dataset-specific settings, and Appendix~\ref{app:prompt_details} provides the prompts used for attribute extraction and summary editing. The released code contains the full preprocessing pipeline and the training configurations.

\item {\bf Open access to data and code}
    \item[] Question: Does the paper provide open access to the data and code, with sufficient instructions to faithfully reproduce the main experimental results, as described in supplemental material?
    \item[] Answer: \answerYes{}
    \item[] Justification: All datasets used (H\&M, MovieLens 20M with the accompanying YouTube trailers mapping, and the Million Song Dataset) are publicly available, and all models used in the pipeline (Qwen3-VL-32B-Instruct, VideoLLaMA~3, MusicFlamingo, Qwen3-30B-A3B-Instruct-2507, BGE-M3, Qwen3.5-27B, Qwen3.5-0.8B, GLM-4-32B-0414, and Gemma-4 12B) have publicly released weights. The CCBR codebase is publicly available at \url{https://github.com/firatoncel/CCBRNeurIPS2026}, including preprocessing scripts, training configurations for all CF backbones, and the controllability evaluation pipeline; the accompanying README provides instructions for reproducing the reported results.
\begin{itemize}
        \item The answer \answerNA{} means that paper does not include experiments requiring code.
        \item Please see the NeurIPS code and data submission guidelines (\url{https://neurips.cc/public/guides/CodeSubmissionPolicy}) for more details.
        \item While we encourage the release of code and data, we understand that this might not be possible, so \answerNo{} is an acceptable answer. Papers cannot be rejected simply for not including code, unless this is central to the contribution (e.g., for a new open-source benchmark).
        \item The instructions should contain the exact command and environment needed to run to reproduce the results. See the NeurIPS code and data submission guidelines (\url{https://neurips.cc/public/guides/CodeSubmissionPolicy}) for more details.
        \item The authors should provide instructions on data access and preparation, including how to access the raw data, preprocessed data, intermediate data, and generated data, etc.
        \item The authors should provide scripts to reproduce all experimental results for the new proposed method and baselines. If only a subset of experiments are reproducible, they should state which ones are omitted from the script and why.
        \item At submission time, to preserve anonymity, the authors should release anonymized versions (if applicable).
        \item Providing as much information as possible in supplemental material (appended to the paper) is recommended, but including URLs to data and code is permitted.
    \end{itemize}

\item {\bf Experimental setting/details}
    \item[] Question: Does the paper specify all the training and test details (e.g., data splits, hyperparameters, how they were chosen, type of optimizer) necessary to understand the results?
    \item[] Answer: \answerYes{}
    \item[] Justification: Section~\ref{sec:experimentalsetup} reports the datasets, the split sizes under strong generalization, and the models used for profile generation and editing. The training objectives are defined in Section~\ref{sec:method} and Appendix~\ref{app:training_objectives}. Appendix~\ref{app:training_setup} reports the hyperparameter search (30 to 50 configurations per backbone and dataset), learning rates, batch sizes, dropout rates, and loss weighting for both training stages, and Appendix~\ref{app:dataset_settings} reports the dataset-specific latent dimensionalities and the H\&M color augmentation. Remaining details, including the optimizer, the InfoNCE temperature, the loss weights, and the selected configuration for each backbone and dataset, are provided in the configuration files of the released code.

\item {\bf Experiment statistical significance}
    \item[] Question: Does the paper report error bars suitably and correctly defined or other appropriate information about the statistical significance of the experiments?
    \item[] Answer: \answerNo{}
    \item[] Justification: We do not report error bars for the recommendation accuracy results in Tables~\ref{tab:results}, \ref{tab:llm_encoder}, and~\ref{tab:beyond_cf}, due to the computational cost of repeating the full pipeline (profile generation, two-stage text-encoder training, and alignment) with multiple seeds for every backbone and dataset. The controllability results in Sections~\ref{sec:controllability}, \ref{sec:loi_unseen}, and~\ref{sec:ml20m_negative} are averages over many per-user, per-attribute interventions, and the same qualitative trends hold across all CF backbones. Latency measurements in Table~\ref{tab:cost} are reported as mean $\pm$ standard deviation over 100 simulated users.

\item {\bf Experiments compute resources}
    \item[] Question: For each experiment, does the paper provide sufficient information on the computer resources (type of compute workers, memory, time of execution) needed to reproduce the experiments?
    \item[] Answer: \answerYes{}
    \item[] Justification: The computational cost analysis in Section~\ref{sec:ml20m_negative} and Table~\ref{tab:cost} report the hardware and execution time of each stage of the pipeline. Offline item verbalization with VideoLLaMA~3 takes $4672.9 \pm 2615.4$ ms per item on a single H100-80GB GPU (about 14.7 GPU-hours for the ML-20M catalog), and serving-time summary generation, summary encoding, and top-20 scoring for a new user are measured on a single A100-80GB GPU.

\item {\bf Code of ethics}
    \item[] Question: Does the research conducted in the paper conform, in every respect, with the NeurIPS Code of Ethics \url{https://neurips.cc/public/EthicsGuidelines}?
    \item[] Answer: \answerYes{}
    \item[] Justification: The work uses publicly available datasets in line with their stated terms of use, does not involve new data collection from human subjects, and does not collect or release personally identifying information. We have reviewed the NeurIPS Code of Ethics and the research conforms to its requirements.

\item {\bf Broader impacts}
    \item[] Question: Does the paper discuss both potential positive societal impacts and negative societal impacts of the work performed?
    \item[] Answer: \answerYes{}
    \item[] Justification: Appendix~\ref{app:broader_impacts} discusses both. 
    
    

\item {\bf Safeguards}
    \item[] Question: Does the paper describe safeguards that have been put in place for responsible release of data or models that have a high risk for misuse (e.g., pre-trained language models, image generators, or scraped datasets)?
    \item[] Answer: \answerNA{}
    \item[] Justification: The paper does not release new large pretrained models, image generators, or scraped datasets. It uses publicly available datasets and publicly released foundation models in their intended modes of use.

\item {\bf Licenses for existing assets}
    \item[] Question: Are the creators or original owners of assets (e.g., code, data, models), used in the paper, properly credited and are the license and terms of use explicitly mentioned and properly respected?
    \item[] Answer: \answerYes{}
    \item[] Justification: All datasets, foundation models, and CF baselines (EASE, EDLAE, Linear-DAE, Multi-DAE, Multi-VAE, MacridVAE, NeuMF, LightGCN) are cited at their first use. Datasets: H\&M Personalized Fashion Recommendations (Kaggle competition data terms), MovieLens 20M (GroupLens usage license, non-commercial research use), the MovieLens 20M YouTube Trailers mapping (CC BY 4.0), and the Million Song Dataset (research terms of use). Models: Qwen3-VL-32B-Instruct, Qwen3-30B-A3B-Instruct-2507, Qwen3.5-27B, Qwen3.5-0.8B, VideoLLaMA~3, and Gemma-4 12B (Apache 2.0); BGE-M3 and GLM-4-32B-0414 (MIT); MusicFlamingo (NVIDIA OneWay Noncommercial License). All assets are used for non-commercial academic research in accordance with these terms.

\item {\bf New assets}
    \item[] Question: Are new assets introduced in the paper well documented and is the documentation provided alongside the assets?
    \item[] Answer: \answerYes{}
    \item[] Justification: The only new asset is the CCBR codebase, released at \url{https://github.com/firatoncel/CCBRNeurIPS2026} together with configuration files for the reported experiments and a README describing data preparation, training, and the controllability evaluation pipeline. No new datasets are released.
\begin{itemize}
        \item The answer \answerNA{} means that the paper does not release new assets.
        \item Researchers should communicate the details of the dataset\slash code\slash model as part of their submissions via structured templates. This includes details about training, license, limitations, etc. 
        \item The paper should discuss whether and how consent was obtained from people whose asset is used.
        \item At submission time, remember to anonymize your assets (if applicable). You can either create an anonymized URL or include an anonymized zip file.
    \end{itemize}

\item {\bf Crowdsourcing and research with human subjects}
    \item[] Question: For crowdsourcing experiments and research with human subjects, does the paper include the full text of instructions given to participants and screenshots, if applicable, as well as details about compensation (if any)? 
    \item[] Answer: \answerNA{}
    \item[] Justification: The paper does not involve crowdsourcing or research with human subjects.

\item {\bf Institutional review board (IRB) approvals or equivalent for research with human subjects}
    \item[] Question: Does the paper describe potential risks incurred by study participants, whether such risks were disclosed to the subjects, and whether Institutional Review Board (IRB) approvals (or an equivalent approval/review based on the requirements of your country or institution) were obtained?
    \item[] Answer: \answerNA{}
    \item[] Justification: The paper does not involve crowdsourcing or research with human subjects, so no IRB approval is required.

\item {\bf Declaration of LLM usage}
    \item[] Question: Does the paper describe the usage of LLMs if it is an important, original, or non-standard component of the core methods in this research? Note that if the LLM is used only for writing, editing, or formatting purposes and does \emph{not} impact the core methodology, scientific rigor, or originality of the research, declaration is not required.
    \item[] Answer: \answerYes{}
    \item[] Justification: LLMs are a core component of the CCBR pipeline and are described in Sections~\ref{sec:method} and~\ref{sec:experimentalsetup}. Modality-specific multimodal foundation models (Qwen3-VL-32B-Instruct for images, VideoLLaMA~3 for videos, and MusicFlamingo for audio) verbalize items, and an instruction-tuned text-only LLM (Qwen3-30B-A3B-Instruct-2507) aggregates item descriptions into user profiles. Qwen3.5-27B is used for attribute extraction and counterfactual summary editing in the controllability experiments, GLM-4-32B-0414 and Gemma-4 12B serve as independent judges of edit locality (Appendix~\ref{app:edit_locality}), and Qwen3.5-0.8B is evaluated as an alternative text encoder (Appendix~\ref{sec:llm_encoder}). Beyond this core methodological role, LLMs were also used for editing (e.g., grammar, spelling, word choice) and for facilitating and running experiments; all LLM-generated outputs and edits were reviewed and verified by the authors.

\end{enumerate}
\end{document}